\documentclass[10pt,twocolumn]{article} 
\usepackage{simpleConference}
\usepackage{times}
\usepackage{graphicx}
\usepackage{amssymb}
\usepackage{url,hyperref}
\usepackage{microtype}
\usepackage{subcaption}
\usepackage{color,soul} 
\usepackage{algorithm}
\usepackage{algpseudocode}
\usepackage{booktabs}

\usepackage{multirow}% http://ctan.org/pkg/multirow
\usepackage{hhline}% http://ctan.org/pkg/hhline
\usepackage{blindtext}
\usepackage{siunitx}
\usepackage{amsmath}
\usepackage{authblk}
\usepackage{siunitx}
\usepackage{amsmath}
\usepackage{amssymb}
\usepackage[section]{placeins}
\usepackage{mathtools}
\usepackage{amsthm}
\usepackage{multicol}
\usepackage{dcolumn}
\usepackage{enumitem}
\usepackage{authblk}
\usepackage[capitalize,noabbrev]{cleveref}
\theoremstyle{plain}

\theoremstyle{definition}

\theoremstyle{remark}

\usepackage[textsize=tiny]{todonotes}

\begin{document}

\title{Modelling spatio-temporal trends of  air pollution in Africa}
\author[1,2]{Paterne Gahungu}
\author[1]{Jean Remy Kubwimana}
\author[2]{Lionel Jean Marie Benjamin Muhimpundu}
\author[3]{Egide Ndamuzi}
%\author[2]{Michael Smith}
%\author[3]{Richard D. Wilkinson}
\affil[1]{African Institute for Mathematical Sciences, Rwanda}
\affil[2]{Institute of Applied Statistics, University of Burundi}
\affil[3]{Ecole Normale Superieur de Bujumbura, Burundi}

\maketitle
\thispagestyle{empty}

\vskip 0.3in

\setlist[itemize]{noitemsep, topsep=0pt}

\setlist[enumerate]{noitemsep, topsep=0pt}

\begin{abstract}
Atmospheric pollution remains one of the major public health threat worldwide with an estimated 7 millions deaths annually. In Africa, rapid urbanization and poor transport infrastructure are worsening  the problem. In this paper, we have analysed spatio-temporal variations of PM2.5 across different geographical regions in Africa. The West African region remains the most affected by the high levels of pollution with a daily average of $\SI{40.856}{\micro\gram}/m^3$ in some cities like Lagos, Abuja and Bamako. In East Africa, Uganda is reporting the highest pollution level with a daily average concentration of $\SI{56.14}{\micro\gram}/m^3$ and $\SI{38.65}{\micro\gram}/m^3$ for Kigali. In countries located in the central region of Africa, the highest daily average concentration of PM2.5 of $\SI{90.075}{\micro\gram}/m^3$ was recorded in N'Djamena. We compare three data driven models in predicting future trends of pollution levels. Neural network is outperforming Gaussian processes and ARIMA models.

\end{abstract}

\section{Introduction }
Air pollution in the form of particulate matter is associated with negative health effects and is considered as the largest contributor  to premature deaths worldwide \cite{kim2015review}\cite{wong2015satellite}\cite{rajak2020short}\cite{shehab2019effects}.  According to the World Health Organization, more than 90 percent of the world's population are exposed to harmful pollutants with levels exceeding up to five times the new guidelines updated in September 2021 by W.H.O \cite{world2021global}.   The major sources of air pollution are attributed to anthropogenic activities like industries, use of highly polluting car for transport, agriculture, and pollution from cooking with fossil fuel. Data on pollution levels and chemical composition are scarce in many low and middle income countries particularly in Africa. This is due to high cost of equipment needed to collect and analyze samples. In most low and middle-income countries, the only data available comes from estimates from satellites. A limited number of research outputs exist in some parts of the world, like the Sub-Saharan African region. The existing literature on air pollution in Africa focus more on the use of low-cost sensors to measure some specific pollutants most likely particulate matter with diameter less than 2.5 micro-meter\cite{okure2022characterization}\cite{raheja2022network} \cite{mcfarlane2021application} \cite{gahungu2022trend}. Some studies have considered short-term and localized campaigns for source apportionment \cite{schwander2014ambient} \cite{kalisa2018characterization}. 

This paper focuses on the spatio-temporal variations of air pollution across Africa and explores predictive models for future trends. The remainder of this work is organized as follows. The second session gives a brief description of the data and models used in this study. The third session presents results for spatio-temporal trends and predictions based on three data-driven models: Auto-Regressive Integrated Moving Average, Neural Network, and Gaussian processes. We conclude the work in session four.

%%%%%%%%%%%%%%%%%%%%%%%%%%%%%%%%%%%%%%%%%%%%%%%%%%%%%%%%%%%%%%%%%
\section{Data and Methods}

\subsection{Data}
\begin{figure}[ht!]
\centering
		% include second image
\caption{Air pollution monitoring stations at US Embassies in Africa}
\includegraphics[width=1.02\linewidth]{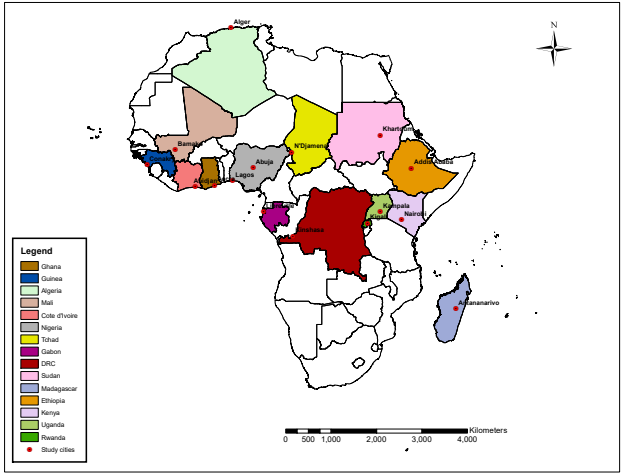}  
\label{africa}
\end{figure}

In this work, we focus on particulate matter with a diameter of less than 2.5 micrometers (PM2.5). It is considered the most dangerous pollutant that affects human health with its ability to easily penetrate the lungs. Reliable measurements come from reference monitors. Reference monitors like the Beta Attenuation Monitors (BAMs) are very expensive. This is the main reason why these monitors are not widely used in many African countries, where the data available comes from low-cost sensors.
The table \textbf{[\ref{africa}]} and  map \textbf{[\ref{africa}]} below describe the African cities studied based on air pollution data collected by the US Embassies across Africa.

\begin{table}[ht!]
\caption{The summary of PM2.5 ($\SI{}{\micro\gram}/m^3$) levels from African cities includes the latitude and longitude of data collection, the time period, and the countries and cities.}
\centering
\resizebox{\columnwidth}{!}{%
\begin{tabular}{ c@{\hspace{2mm}} c @{\hspace{1.5\tabcolsep}} c @{\hspace{1.5\tabcolsep}} c @{\hspace{2mm}}c }
\toprule
\textbf{African Cities}              & \textbf{Country} & \textbf{Period} &   \textbf{Latitude} &   \textbf{Longitude }\\
\toprule
Kigali & Rwanda & Feb - June 2022 & 1.9441° S  & 30.0619° E   \\
Conakry & Guinea &  Jan 2020 - June 2022&  9.6412° N & 13.5784° W     \\
Bamako           & Mali&   Oct 2019 - Jan 2022 & 12.6392° N  &    8.0029° W    \\
Lagos     & Nigeria &  Feb 2021 - June 2022  & 6.5244° N  &   3.3792° E    \\
Abuja&    Nigeria &  Feb 2021 - June 2022&9.0765° N   &   7.3986° E  \\
Libreville&  Gabon & Apr 2021- Mar 2022 & 0.4162° N  &   9.4673° E    \\
Algiers&  Algeria & Feb 2019 - June 2022& 36.7538° N   & 3.0588° E     \\
Accra&  Ghana & Jan 2020 - June 2022 &5.6037° N   &  0.1870° W    \\
Kinshasa&  DRC &  Mar - June 2022  &4.4419° S  &   15.2663° E   \\
Kampala&  Uganda &  Feb 2017 - June 2022 &0.3476° N &   32.5825° E     \\
Nairobi&  Kenya  & Mar 2021 - June 2022&  1.2921° S&   36.8219° E      \\
N'Djamena& Tchad & May 2020 - June 2022 & 12.1348° N &     15.0557° E  \\
Khartoum& Sudan &  Jan 2020 - June 2022 & 15.5007° N  &  32.5599° E     \\
Addis Ababa& Ethiopia  & Aug 2016 - June 2022  & 8.9806° N &  38.7578° E     \\
Antananarivo&  Madagascar &  Jan 2020 - June 2022 & 18.8792° S  &    47.5079° E    \\
Abidjan& C\^{o}te d'Ivoire &  Feb 2020 - June 2022 & 5.3600° N  &    4.0083° W    \\
\bottomrule
\end{tabular}
}
\label{table}
\end{table}
\FloatBarrier
\FloatBarrier
\subsection{Methods}
We model the data generation process with the  Auto-Regressive Integrated Moving Average (ARIMA) model and two universal function approximators: Neural networks and Gaussian processes.
\subsection*{Auto-Regressive Integrated Moving Average}
We model  $y_t$, the observation at time $t$ as a function of historical values $y_{0}, \cdots, y_{t-1}$. 

\begin{eqnarray}
y_t=f(y_0,\cdots, y_{t-1}).
\end{eqnarray}
where
\begin{align*}
f= \alpha+ \beta_1 y_{t-1}+\beta_2 y_{t-2}+\cdots+ \beta_{p} y_{t-p}
+ \epsilon_t \\ \nonumber +\mu+W_t+\theta_1W_{t-1}+\theta_2W_{t-2}+\cdots+\theta_qW_{t-q} \nonumber,
\end{align*}
where the model coefficients are $\beta_1, \cdots, \beta_p$, and $\theta_1, \cdots, \theta_q$, $p$ and $q$ are non-negative integers at any lag, $\alpha$ and $\mu$ are the model intercepts, and $\epsilon_t$ is an error term such that $\epsilon_t \sim \mathcal{N} (0,1)$, $\mu$ is the long run average, $W_t$ is the shock for the process such that $W_t= \sigma \times \epsilon_t$ where $\sigma$ is the conditional standard deviation. 
% where $W_t$ is white noise, 

\subsection*{Neural networks}
We denote the observations at time $j$ by $x_j$ with $j=1, \cdots, n$. Here $n$ is size of the dataset. An approximation of the data generation process using a neural networks can be expressed as follows:

\begin{eqnarray}
y=f\left (\sum_{j=1}^{n-1} W_j x_j+b \right),
\end{eqnarray}
where $W$ are weights, $b$ is the bias and $f$ is the activation function. The following activation functions are commonly used.

\begin{itemize}
    \item Rectified Linear Unit (ReLU).
    \begin{eqnarray}
    f(.)=\max(0,.).
    \end{eqnarray}
    \item Sigmoid function.
    \begin{eqnarray}
    f(.)=\frac{1}{1+\exp(-.)}.
    \end{eqnarray}
    \item Hyperbolic tangent.
    \begin{eqnarray}
    f(.)=\tanh(.)=\frac{\exp(.)-\exp(-.)}{\exp(.)+\exp(-.)}.
    \end{eqnarray}
\end{itemize}

\subsection*{Gaussian processes}
Gaussian processes are considered as infinite-dimensional normal distributions with a mean and a covariance functions denoted by $m(.)$ and $k(.,.')$ respectively \cite{rasmussen2003gaussian}.

\begin{eqnarray}
f(.)\sim GP(m(.),k(.,.'))
\end{eqnarray}
The following covariance functions are commonly used:
\begin{itemize}
    \item Radial Basis Function (RBF).
    \begin{eqnarray}
    k(.,.')=\sigma^2\exp(-{\frac{||x-x'||^2}{2l^2}})
    \end{eqnarray}
    \item Linear function.
    \begin{eqnarray}
    k(.,.')=\sum_{i=1}^{N} \sigma_i^2 x_i x_{i}^{'}
    \end{eqnarray}
\end{itemize}

\section{Results and Discussion}
\subsection{Trend analysis}
Air pollution and its emission in African cities differ based on the country's population, industrialization, urbanization, economic status, emission sources, physical regions, and meteorology. The trend analysis of PM2.5 levels included sixteen cities from across Africa. The analysis shows that the mass concentrations of PM2.5 in those cities are currently increasing, negatively impacting people's lives by causing respiratory and cardiovascular diseases. The hourly, daily and seasonal analyses were made based on the  cities' locations. 

\begin{figure}[ht!]
    \centering
		\caption{Hourly PM2.5 concentrations in East Africa}
		% include second image
		\includegraphics[width=1.02\linewidth]{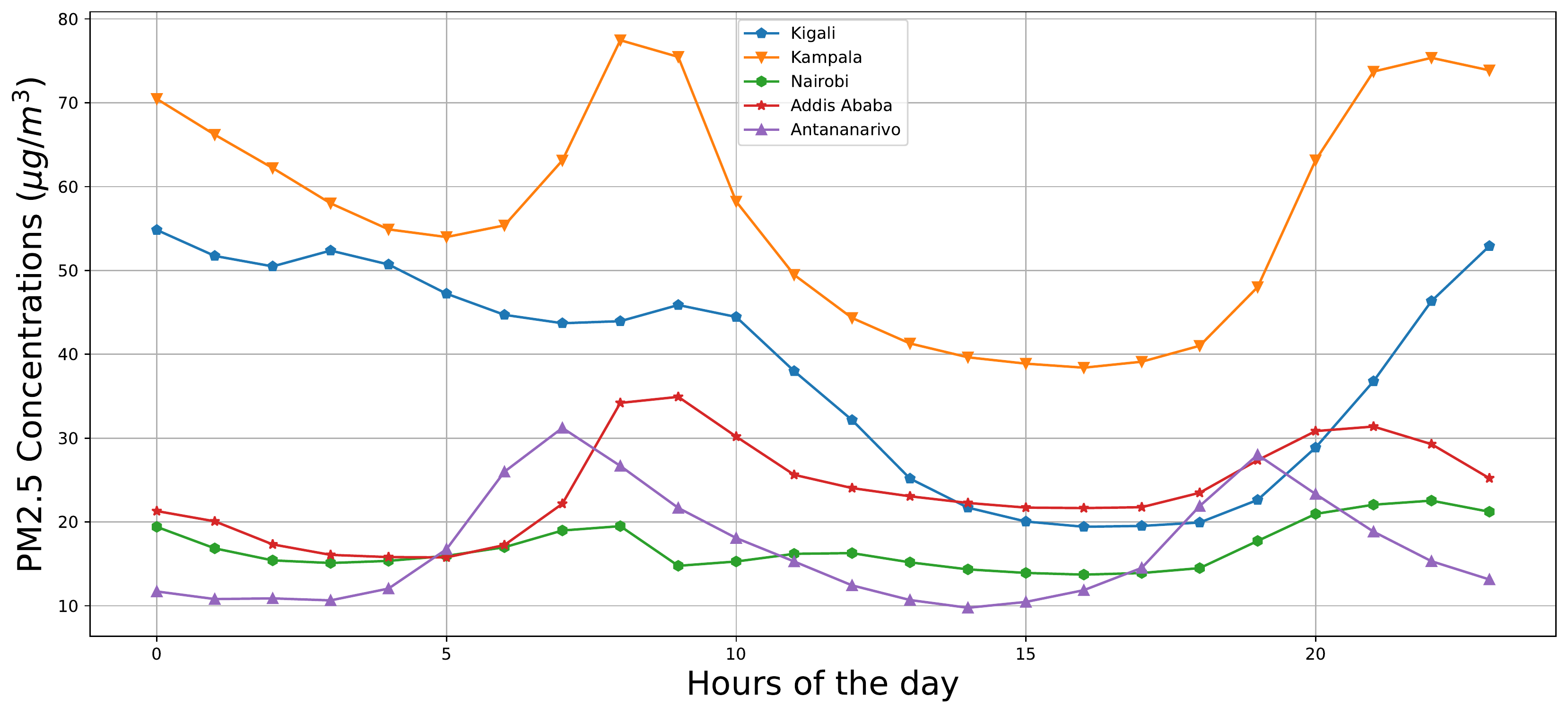}  
		\label{fig12}
\end{figure}

The figure \textbf{[\ref{fig12}]} shows hourly mean concentrations of PM2.5 in the countries from East Africa. From the figure, the highest hourly mean concentration of PM2.5 in the morning hours was observed in Kampala at 8:00 a.m. with the values of $\SI{77.46}{\micro\gram}/m^3$, and the trend started to decline in the period of 8:00 a.m. – 3:00 p.m. After 3:00 p.m., the hourly trend begins to rise as many workers with vehicles return home and emissions from biomass burning as a source of energy at home were suggested to contribute. 

\vspace{0.9mm}
Kigali ranks second in the morning rush hours when workers are on their way to work, with high levels of hourly mean concentration at 9:00 a.m. PM2.5 levels in Kigali begin to rise again between 4:00 p.m. and late at night, with the highest hourly mean concentration of $\SI{54.82}{\micro\gram}/m^3$ recorded at 0:00 a.m. These differences in PM2.5 levels in Kigali are thought to be caused by traffic, industries, and combustion activities. Other developing cities, such as Nairobi, Antananarivo and Addis Ababa have PM2.5 concentrations that are less than the ones for Kigali and Kampala. Though the mass concentrations of PM2.5 in Nairobi are not as high as in other cities such as Kampala, Kigali, and Addis Ababa, according to \textbf{\cite{gaita2014source}}, mineral dust and traffic-related factors account for  $74\%$ of annual PM2.5 concentrations of the entire country.
 
\textbf{\cite{kinney2011traffic}} In their findings have shown that PM2.5 concentrations in Nairobi, particularly in the street and areas adjacent to a city road, have exceeded WHO guidelines of $\SI{25 }{\micro\gram}/m^3$ as a 24-hour mean. The East African traffic network is expanding on a daily basis, resulting in high levels of traffic activity; however, the majorities of traffic in these cities are old and in poor condition. These issues contribute to the respiratory problems that pedestrians may experience as a result of the improved road networks, particularly during morning and evening rush hours.
%%%%%%%%%%%%%%%%%%%%%%%%%%%%%%%%%%%%%%%%%%%%%%%%%%%%%%%%%%%%%%%%%%%%%%%%%%%%%%%%%%%%%%%%%%%%%%%%%%%%%%%%%%%%%%%%%%%%%%%%%%%%%%%%%%%%%%%%%%%%%%%%%%%%%%%%%%%%%%
\FloatBarrier
%%%%%%%%%%%%%%%%%%%%%%%%%%%%%%%%%%%%%%%%%%%%%%%%%%%%%%%%%%%%%%%%%%%%%%%%%%%%%%%%%%%%%%%%%%%%%%%%%%%%%%%%%%%%%%%%%%%%%%%%%%%%%%%%%%%%%%%%%%%%%%%%%%%%%%%%%%%%%%%%%%%%%%%%%%%%%%%%%%%%%%%%%%%%%%%%%%%%%%%%%%%%%%%%%%%%%%%%
\textbf{\cite{leon2021pm}} Environmental degradation has put pressure on West African countries as a result of rapid population growth, urbanization, and a growing economy. Small particle concentrations in West Africa are frequently much higher than the WHO recommended limits \textbf{\cite{  world2021global}}. These fine particles are caused by human activities such as charcoal fires, waste combustion in cities, and Savannah fires. Other particles originate in North Africa as a result of the wind blowing dust from the Sahara desert.

\begin{figure}[ht!]
    \centering
		\caption{Hourly PM2.5 concentrations in West Africa}
		% include second image
		\includegraphics[width=1.02\linewidth]{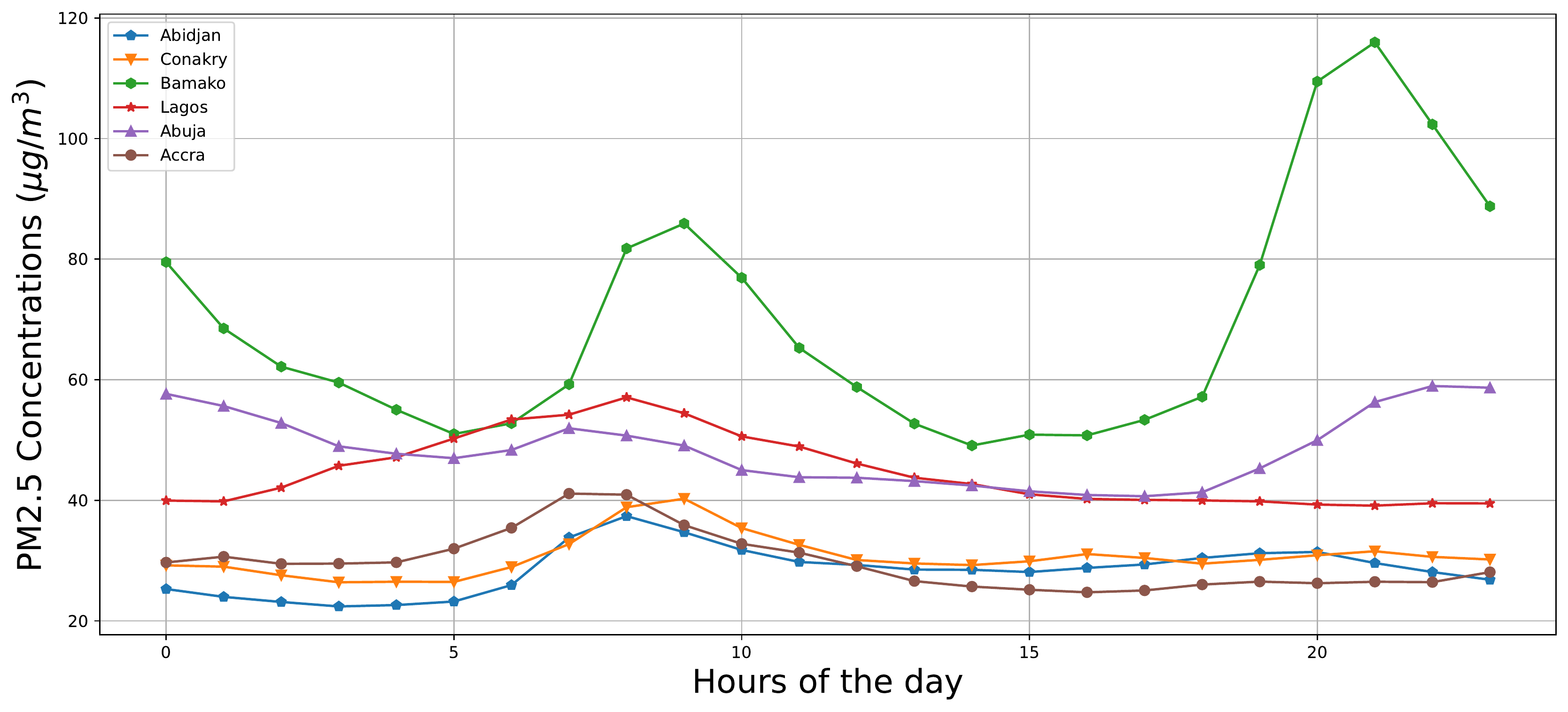}  
		\label{fig13}
\end{figure}

The figure \textbf{[\ref{fig13}]} depicts the hourly mean mass concentrations of PM2.5 in West African cities. According to the figure Bamako experiences the highest hourly mean PM2.5 concentrations due to its location. Emissions in Bamako have been linked to traffic, biomass burning, dust, and desert dust events. All of these factors are thought to have contributed to this city having the highest observable PM2.5 levels of any West African city studied. 

Higher morning peak time averages in Bamako was observed at 9:00 a.m., with values of $\SI{85.89}{\micro\gram}/m^3$, and during the evening hours, the trend began to rise from 3:00 p.m. to 9:00 p.m. where the highest values of $\SI{115.95 }{\micro\gram}/m^3$ was recorded at 9:00 p.m. This has resulted in this city having the highest PM2.5 emissions among the remaining West African cities considered in the study.

\textbf{\cite{ezeh2017elemental}} Nigeria, as one of Africa's fastest developing countries, is being affected by the rise in mass concentrations of PM2.5 caused by human activities and industrialization in cities like Lagos and Abuja. According to the World Bank \textbf{\cite{ WorldBank2020lagos}}, Nigeria had the highest number of premature deaths due to ambient PM2.5, particularly in the cities mentioned above. Road transport, industrial emissions, and power generation are the primary sources of air pollution in Nigeria, though more research is needed to determine the contribution of each sector separately. \textbf{\cite{croitoru2020cost}} Because some cars in Nigeria are older than five years, the quality of imported fuels (diesel and gasoline), and the unlimited means of transportation in cities, road transport contributes much more than other sources. All of these factors contributed to Abuja and Lagos having the highest mass concentrations of PM2.5 in the morning rush hours when compared to other remaining cities such as Accra, Conakry, and Bamako, particularly between the hours of 7:00 a.m. and 9:00 a.m. Between 6:00 p.m. and 10:00 p.m. in Abuja, PM2.5 emissions begin to rise due to mineral dust, vehicle exhaust, industrial emissions, and the use of firewood as a cooking fuel. 

\begin{figure}[ht!]
    \centering
		\caption{Hourly PM2.5 concentrations in Central Africa}
		% include second image
		\includegraphics[width=1.02\linewidth]{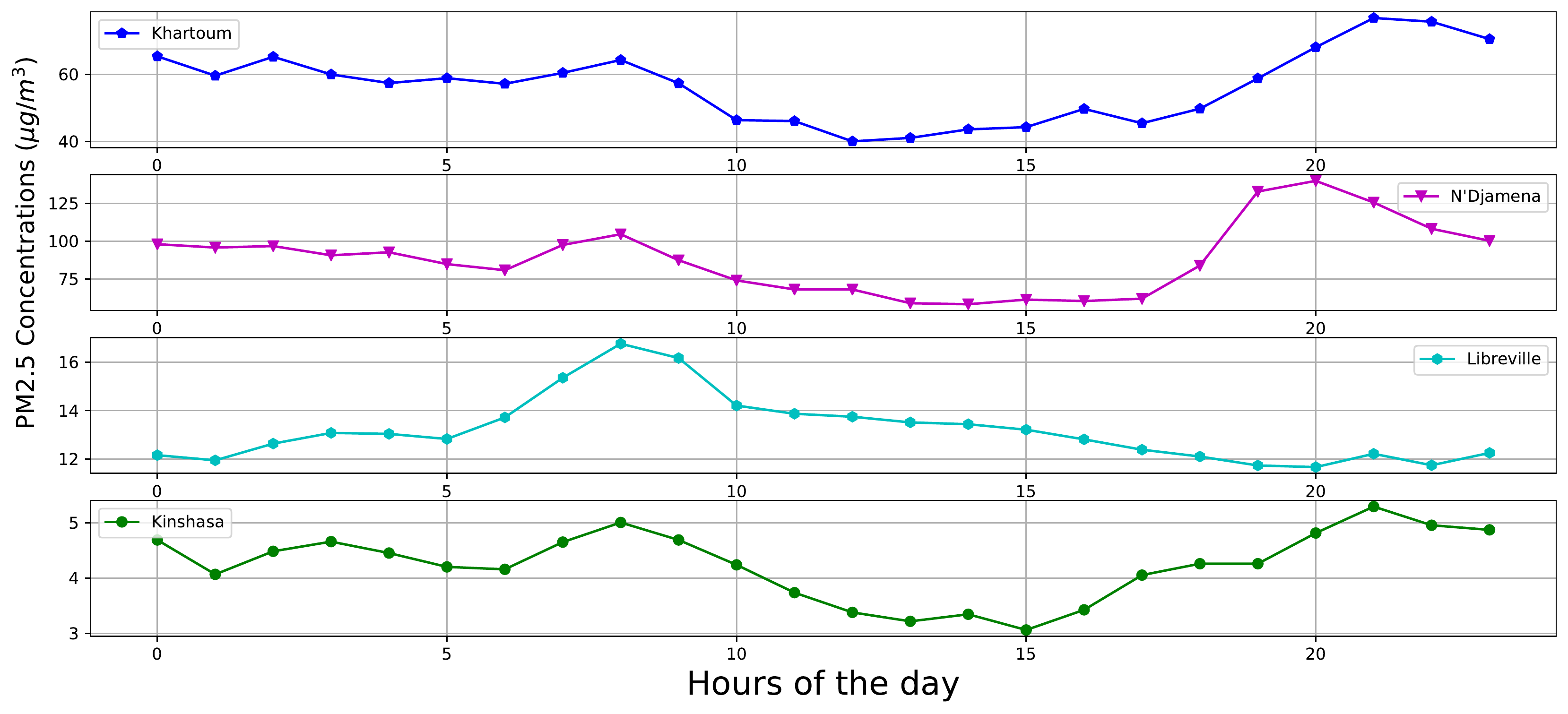}  
		\label{fig11}
\end{figure}

\textbf{\cite{weinstein2010characterization}} In Guinea, particularly in Conakry, some of the proposed air pollution emissions are caused by unregulated combustion and processing emissions from industrial sites, unregulated emissions from gasoline vehicles and residential wood burning. According to \textbf{\cite{weinstein2010characterization}}, cement manufacturing plants, electric power plants, brick manufacturing operations, steel smelters, and medical waste incinerators are among the industries that are thought to contribute to the mass concentration of PM2.5 in Conakry. Conakry has low hourly emissions when compared to other cities such as Bamako, Lagos, and Abuja. At 9:00 a.m., the highest hourly mean mass concentration of PM2.5 was recorded, with a value of $\SI{40.25 }{\micro\gram}/m^3$ whereas the highest hourly mean concentrations of PM2.5 in the morning hours in Bamako, Lagos, and Abuja are $\SI{85.89 }{\micro\gram}/m^3$ at 9:00 a.m., $\SI{57.06 }{\micro\gram}/m^3$ at 8:00 a.m., and $\SI{51.94 }{\micro\gram}/m^3$ at 7:00 a.m respectively. Other cities, such as Abidjan and Accra, which are close to the Atlantic Ocean, have emission rates similar to Conakry, as shown by the hourly trends in the figure \textbf{[\ref{fig13}]}.
\FloatBarrier
%%%%%%%%%%%%%%%%%%%%%%%%%%%%%%%%%%%%%%%%%%%%%%%%%%%%%%%%%%%%%%%%%%%%%%%%%%%%%%%%%%%%%%%%%%%%%%%%%%%%%%%%%%%%%%%%%%%%%%%%%%%%%%%%%%%%%%%%%%%%%%%%%%%%%%%%%%%%%%%%
The hourly mean variation of PM2.5 in figure \textbf{[\ref{fig11}]} compares African cities from central region. According to the graph \textbf{[\ref{fig11}]}, the highest hourly mean mass concentrations of PM2.5 were observed in N'Djamena and Khartoum. These two cities are close to the Sahara desert, which influences PM2.5 emissions in neighboring countries. 
Like in other cities, in the morning hours from 6:00 am to 9:00 am in N'Djamena, the emission has increased up to $\SI{104.53}{\micro\gram}/m^3$ at 8:00 am, while in the evening, the highest hourly mean concentration of $\SI{139.83}{\micro\gram}/m^3$ was observed at 8:00 pm. In Khartoum, the highest peak of $\SI{64.36}{\micro\gram}/m^3$ was observed at 8:00 a.m., and levels of PM2.5 decreased after 8:00 a.m. until noon. The levels gradually increased at a slow rate during the afternoon hours until the evening. The highest peak in the evening and late night was at 9:00 a.m., with a mass concentration of $\SI{76.94}{\micro\gram}/m^3$.

Though PM2.5 concentrations in Kinshasa and Libreville in figure \textbf{[\ref{fig11}]} are not as high as in other cities, the sources of PM2.5 emissions include road transportation, poorly maintained vehicles, smoke from open-air barbeques, burning trash, and non-standard fuels such as gasoline and diesel \textbf{\cite{mcfarlane2021first}}. Gabon as a developing country, it has policies in place to control the quality of imported gasoline and diesel, but no clear policy in place to control vehicle emission standards and air quality regulations. 

\begin{figure}[ht!]
    \centering
		\caption{Hourly PM2.5 concentrations in North Africa}
		% include second image
		\includegraphics[width=1.02\linewidth]{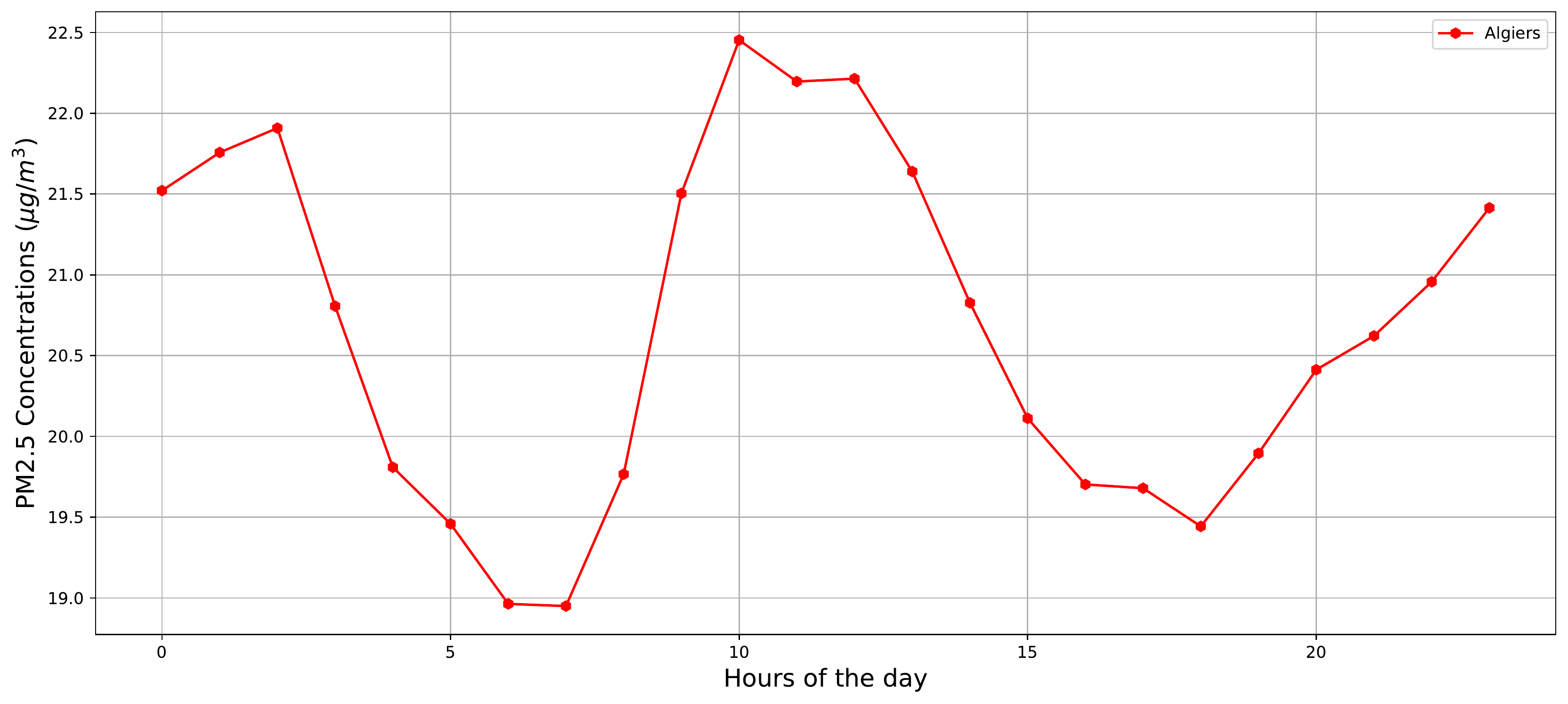}  
		\label{fig14}
\end{figure}

\FloatBarrier
%%%%%%%%%%%%%%%%%%%%%%%%%%%%%%%%%%%%%%%%%%%%%%%%%%%%%%%%%%%%%%%%%%%%%%%%%%%%%%%%%%%%%%%%%%%%%%%%%%%%%%%%%%%%%%%%%%%%%%%%%%%%%%%%%
\textbf{\cite{belarbi2020road}} Fuel combustion, power generation, and industrial facilities are some of the primary sources of PM2.5 in Algiers. Residential fireplaces and wood burning as a source of energy at home are two other sources that have increased emissions. This rise in PM2.5 mass concentrations in Algiers is linked to various cases of mortality and morbidity throughout the country. From figure \textbf{[\ref{fig14}]}, the highest and lowest hourly mean mass concentration of PM2.5 of $\SI{22.45}{\micro\gram}/m^3$ and $\SI{18.94}{\micro\gram}/m^3$ was observed at 10:00 a.m. and 7:00 a.m. respectively. The increase in PM2.5 mass concentrations in Algiers from 7:00 a.m. to 10:00 a.m. is associated with an increase in traffic activities, particularly traffic movements as people commute to work.  PM2.5 levels dropped from 10:00 a.m. throughout the afternoon hours to 19.44/m3 at 6:00 p.m. Then, from 7:00 a.m. until late, the levels of PM2.5 in Algiers increased, reaching $\SI{21.908}{\micro\gram}/m^3$ at 2:00 a.m.

\begin{figure}[ht!]
    \centering
		\caption{Daily PM2.5 concentrations in East Africa}
		% include second image
		\includegraphics[width=1.02\linewidth]{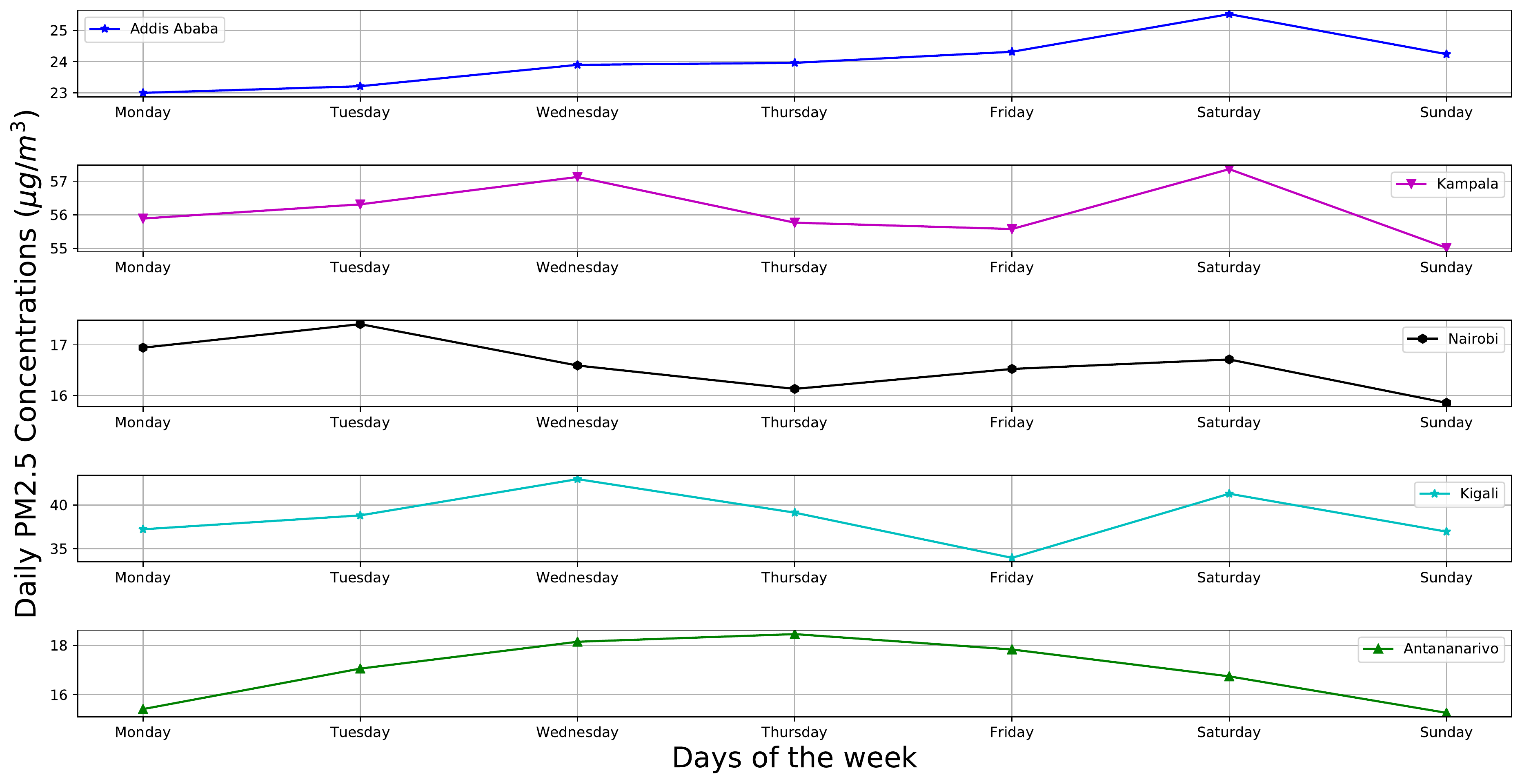}  
		\label{fig2}
\end{figure}

\begin{figure}[ht!]
    \centering
		\caption{Daily PM2.5 concentrations in West Africa}
		% include second image
		\includegraphics[width=1.02\linewidth]{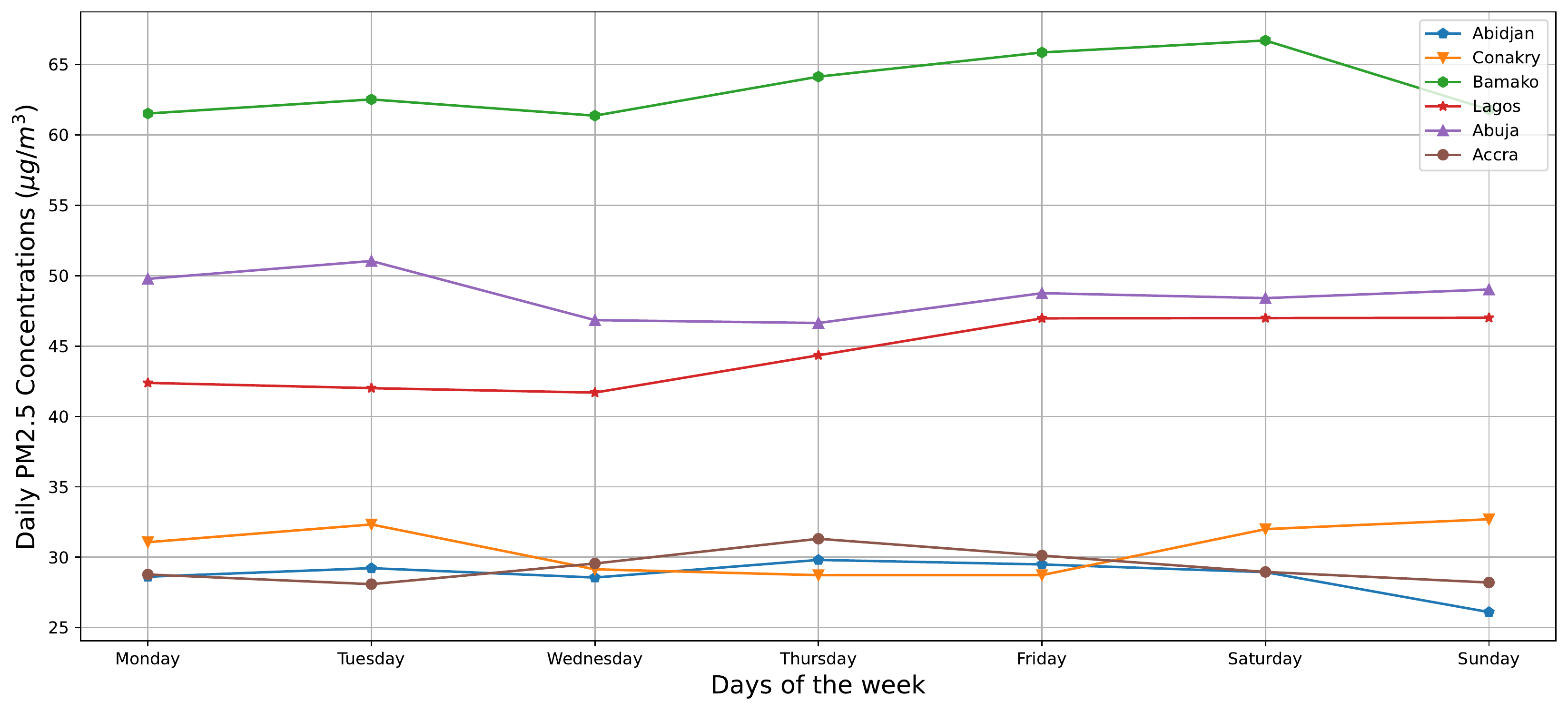}  
		\label{fig3}
\end{figure}

\begin{figure}[ht!]
    \centering
		\caption{Daily PM2.5 concentrations in Central Africa}
		% include second image
		\includegraphics[width=1.02\linewidth]{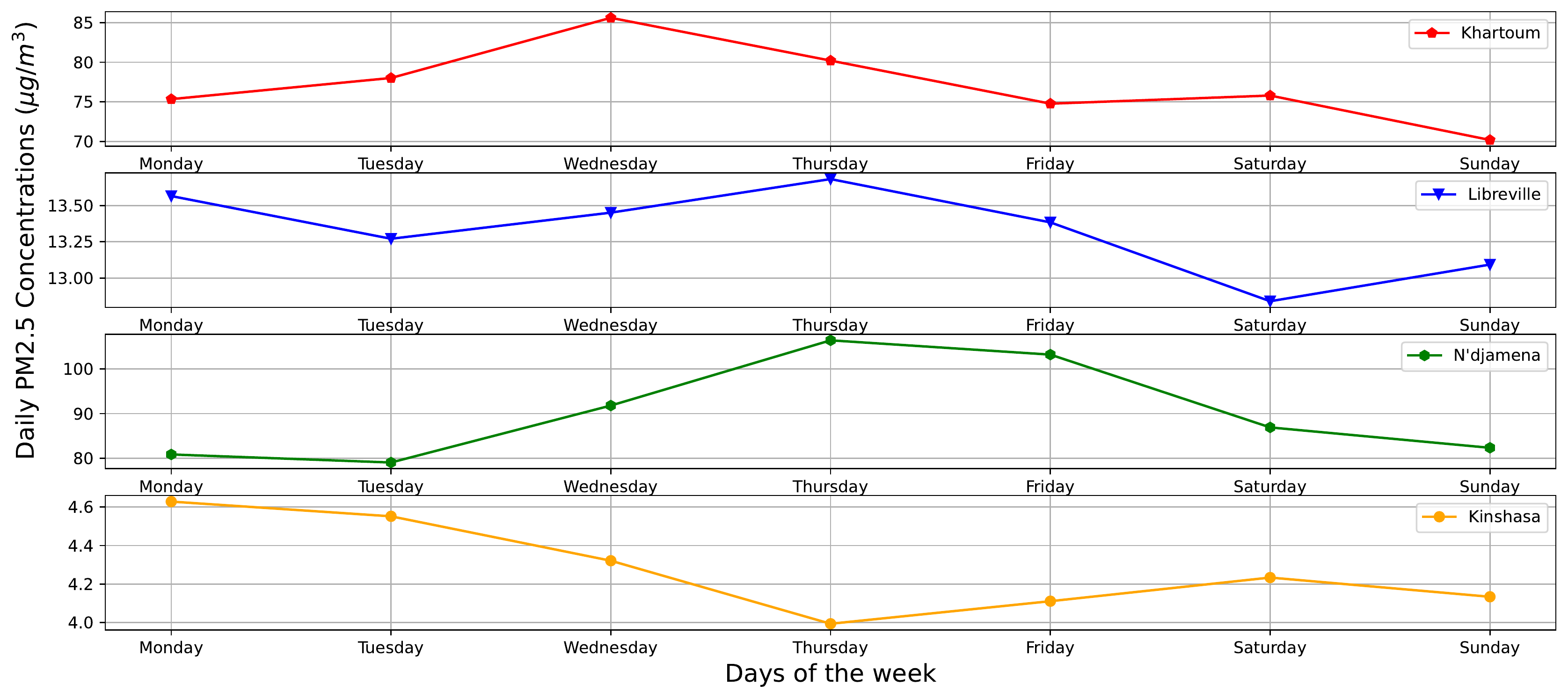}  
		\label{fig1}
\end{figure}

\begin{figure}[ht!]
    \centering
		\caption{Daily PM2.5 concentrations in North Africa}
		% include second image
		\includegraphics[width=1.02\linewidth]{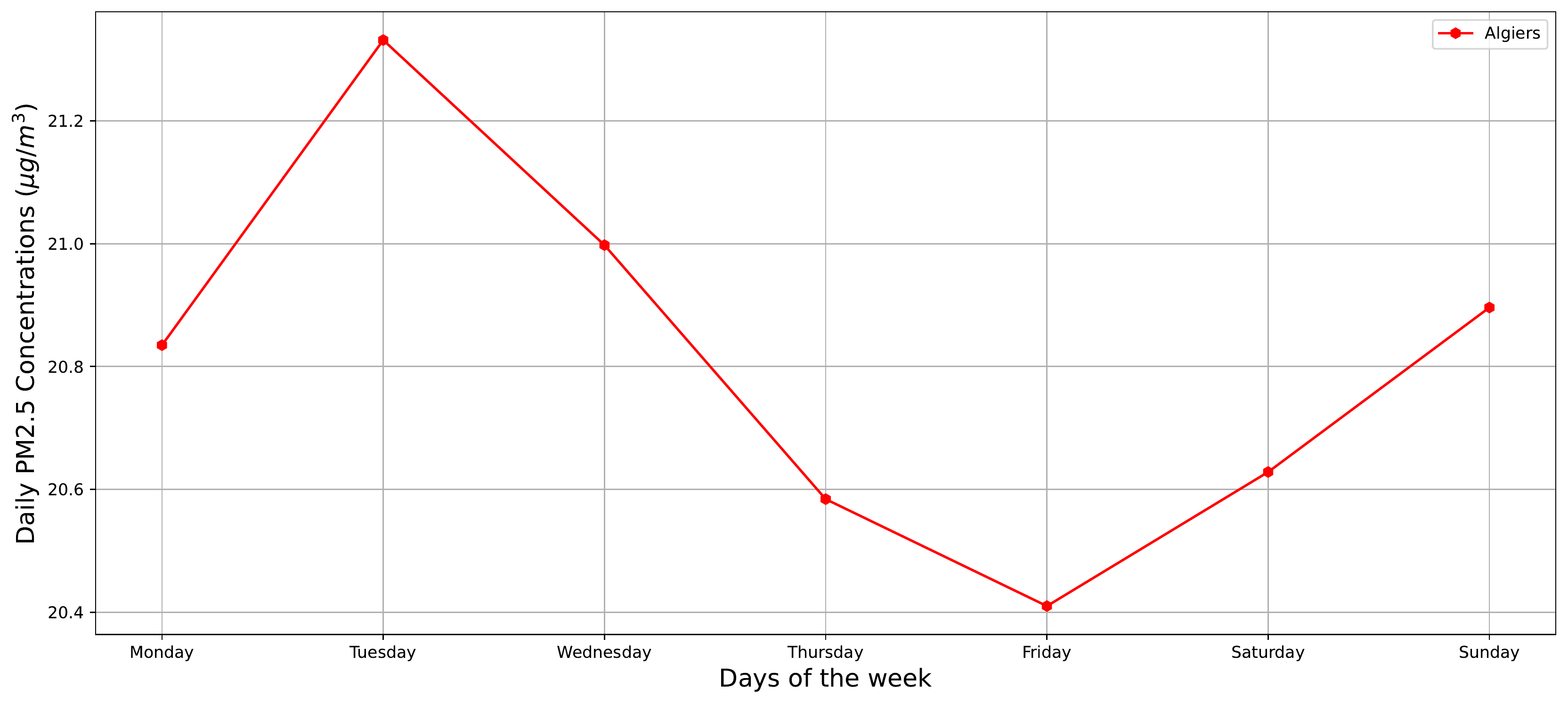}  
		\label{fig4}
\end{figure}

\begin{figure}[ht!]
    \centering
		\caption{Monthly PM2.5 concentrations in East Africa}
		% include second image
		\includegraphics[width=1.02\linewidth]{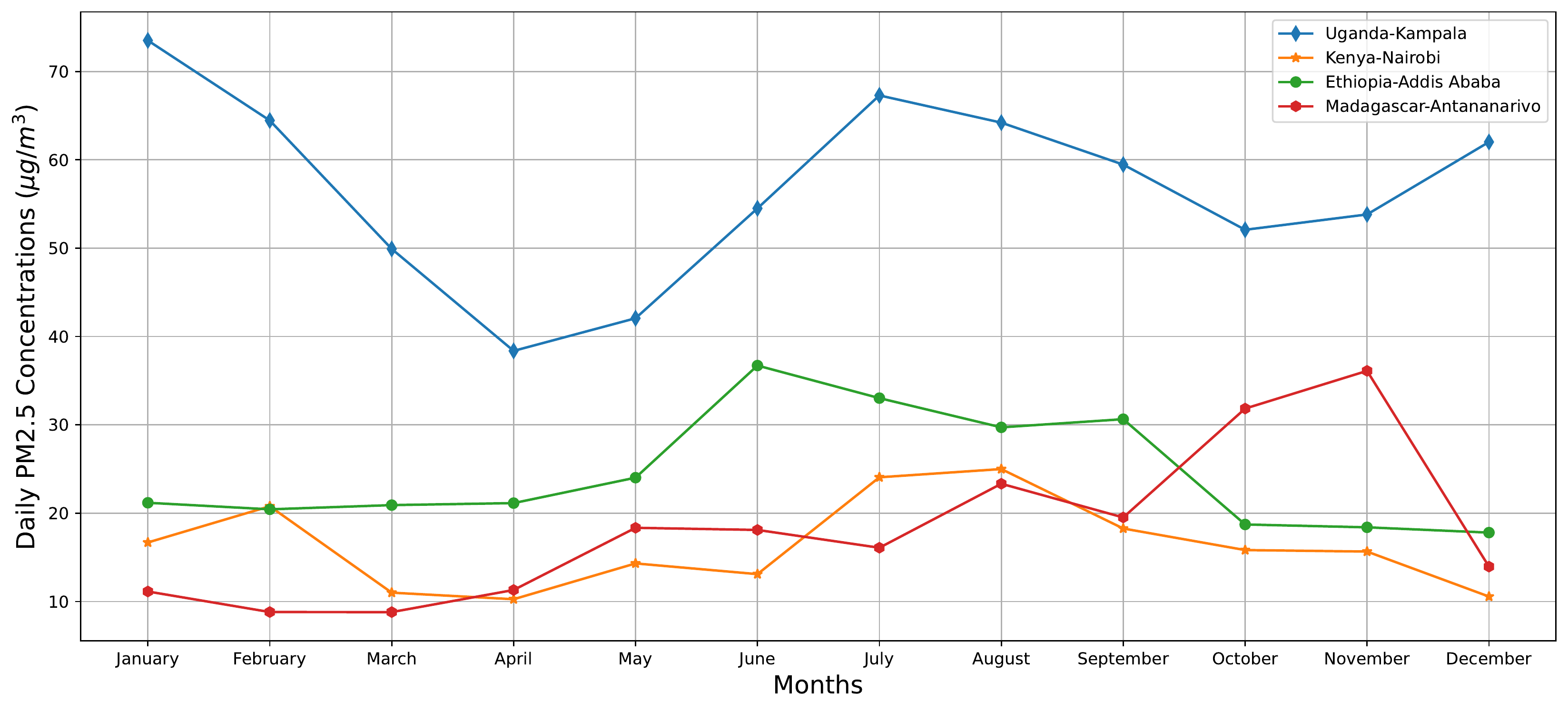}  
		\label{fig:sub-first_monthly}
\end{figure}

\FloatBarrier
%%%%%%%%%%%%%%%%%%%%%%%%%%%%%%%%%%%%%%%%%%%%%%%%%%%%%%%%%%%%%%%%%%%%%%%%%%%%%%%%%%%%%%%%%%%%%%%%%%%%%%%%%%%%%%%%%%%%%%%%%%%%%%%%%%%%%%%%%%%%%%%%%%%%%%%%%%%%%%%%%%%%%%%%%%%%%%%%%%%%%%%%%%%%%%%%
The daily trend analysis in figure \textbf{[\ref{fig2}]} shows that weekdays (Monday through Friday) have high PM2.5 concentrations during the weekends (Saturday and Sunday) in some cities. These differences in emission during the weekdays and weekends are linked to the air pollution resulted from human and traffic activities in east Africa.  Human activities have influenced the rate of emissions during the week, and the decrease on weekends is due to minimal activities. In Kigali, Nairobi, and Antananarivo, the highest peak was observed on weekdays rather than weekends, whereas weekends had higher levels of PM2.5 than weekdays in Kampala and Addis Ababa. The lower levels of PM2.5 in Antananarivo on weekends compared to weekdays were caused by a decrease in traffic activity. The daily variation of PM2.5 concentration in Kampala, Kigali, and Nairobi shows that the highest levels of PM2.5 were observed on Saturday, Wednesday, and Tuesday, with values of $\SI{57.35}{\micro\gram}/m^3$, $\SI{42.97}{\micro\gram}/m^3$, and $\SI{17.40}{\micro\gram}/m^3$, respectively. This implies that because these cities are among the most populated in East Africa, the daily mass concentration of PM2.5 varies due to daily activities. In Addis Ababa, weekday emissions were lower than on weekends, with Saturday having the highest daily mass concentration of $\SI{25.51}{\micro\gram}/m^3$.
\FloatBarrier
%%%%%%%%%%%%%%%%%%%%%%%%%%%%%%%%%%%%%%%%%%%%%%%%%%%%%%%%%%%%%%%%%%%%%%%%%%%%%%%%%%%%%%%%%%%%%%%%%%%%%%%%%%%%%%%%%%%%%%%%%%%%%%%%%%%%%%%%%%%%%%%%%%%%%%%%%%%%%%%%%%%%%%%%%%%%%%%%%%%%%%%%%%%%%%%%%%%%%%%%%%%%%%%%%%%%%%%%%%%%%%
The figure \textbf{[\ref{fig3}]} shows the weekly average PM2.5 for cities in West African countries. The mass concentration of PM2.5 is higher in cities such as Bamako, Lagos, and Abuja than in Accra, Abidjan, and Conakry. In these coastal cities, PM2.5 levels are lower than in other cities, with the exception of Lagos, Africa's most populous city, where daily emissions that pollute the air are high.
\FloatBarrier
%%%%%%%%%%%%%%%%%%%%%%%%%%%%%%%%%%%%%%%%%%%%%%%%%%%%%%%%%%%%%%%%%%%%%%%%%%%%%%%%%%%%%%%%%%%%%%%%%%%%%%%%%%%%%%%%%%%%%%%%%%%%%%%%%%%%%%%%%%%%%%%%%%%%%%%%%%%%%%%%%%%%%%%%%%%%%%%%%%%%%%%%%%%%%%%%%%%%%%%%%%%%%%%%%%%%%%%%%%%%%%
According to figure \textbf{[\ref{fig1}]}, the concentration of PM2.5 was higher on weekdays than on weekends in Khartoum and N'Djamena. The average mass concentration of PM2.5 in Khartoum was $\SI{78.78}{\micro\gram}/m^3$ during the week and $\SI{72.979}{\micro\gram}/m^3$ at the weekend. The average PM2.5 level in N'Djamena was $\SI{92.26}{\micro\gram}/m^3$ during the week and $\SI{84.64}{\micro\gram}/m^3$ on weekends. When these results are compared, weekday emissions are more significant than weekends because they are the days with the most traffic and the influence of agricultural and construction activities is visible on weekdays. Although daily emissions are significant in Khartoum and N'Djamena, they are not the same in Libreville and Kinshasa. The highest mass concentration in PM2.5 was $\SI{13.68}{\micro\gram}/m^3$ in Libreville on Thursday, and $\SI{4.62}{\micro\gram}/m^3$ in Kinshasa on Monday. The daily emissions in these cities were not as high as in Khartoum and N'Djamena.
\FloatBarrier
%%%%%%%%%%%%%%%%%%%%%%%%%%%%%%%%%%%%%%%%%%%%%%%%%%%%%%%%%%%%%%%%%%%%%%%%%%%%%%%%%%%%%%%%%%%%%%%%%%%%%%%%%%%%%%%%%%%%%%%%%%%%%%%%%%%%%%%%%%%%%%%%%%%%%%%%%%%%%%%%%%%%%%%%%%%%%%%%%%%%%%%%%%%%%%%%%%%%%%%%%%%%%%%%%%%%%%%%%%%%%%
Figure \textbf{[\ref{fig4}]} shows the daily variation of PM2.5 levels in Algiers. The overall weekdays mean concentration is $\SI{20.83}{\micro\gram}/m^3$ and $\SI{20.76}{\micro\gram}/m^3$ on weekends. This indicates that emissions were higher during the week than on weekends. The highest value of $\SI{21.33}{\micro\gram}/m^3$ PM2.5 was observed on Monday, while the lowest value of $\SI{20.409}{\micro\gram}/m^3$ was observed on Friday.
\FloatBarrier
%%%%%%%%%%%%%%%%%%%%%%%%%%%%%%%%%%%%%%%%%%%%%%%%%%%%%%%%%%%%%%%%%%%%%%%%%%%%%%%%%%%%%%%%%%%%%%%%%%%%%%%%%%%%%%%%%%%%%%%%%%%%%%%%%%%%%%%%%%%%%%%%%%%%%%%%%%%%%%%%%%
The figure \textbf{[\ref{fig:sub-first_monthly}]} depicts the seasonal variation of PM2.5 concentrations in East African cities. During the dry season, mineral dust is one of the suggested contributors to the increase in particulate matter due to unpaved road surfaces, especially when the wind blows dust. During the dry season, Savannah fires are associated with carbonaceous aerosols detected in this region of East Africa.

In figure \textbf{[\ref{fig:sub-first_monthly}]} seasonal variations in PM2.5 levels were higher in Kampala and Addis Ababa than in other East African cities such as Nairobi and Antananarivo. The highest measured PM2.5 levels in Kampala during the dry season were observed in July, and as the country entered the short rainy season, emissions began to decline until the start of the short dry season. The monthly mean concentration of PM2.5 in the dry season is $\SI{67.29}{\micro\gram}/m^3$ in July and $\SI{73.51}{\micro\gram}/m^3$ in January whereas the lowest values observed in April and May were $\SI{38.37}{\micro\gram}/m^3$ and $\SI{42.06}{\micro\gram}/m^3$, respectively. Similarly, seasonal variation was observed in Addis Ababa, with PM2.5 averaging $\SI{36.71}{\micro\gram}/m^3$ during the dry season and $\SI{21.17}{\micro\gram}/m^3$ during the rainy season.

There are only two seasons in Madagascar: a hot, rainy season from November to April and a cooler, dry season from May to October. The highest level of PM2.5 in November was $\SI{36.11}{\micro\gram}/m^3$ during the hot, rainy season, while it was $\SI{31.84}{\micro\gram}/m^3$ during the other season. Nairobi appears to have lower PM2.5 levels than other cities. This implies that the factors that would be expected to cause seasonal variations (mineral dust suspension, open-air waste burning, and agricultural burning) have emitted fewer pollutants than in other cities. PM2.5 concentrations in Nairobi are higher in July and August than in other months.
\FloatBarrier
%%%%%%%%%%%%%%%%%%%%%%%%%%%%%%%%%%%%%%%%%%%%%%%%%%%%%%%%%%%%%%%%%%%%%%%%%%%%%%%%%%%%%%%%%%%%%%%%%%%%%%%%%%%%%%%%%%%%%%%%%%%%%%%%%%%%%%%%%%%%%%%%%%%%%%%%%%%%%%%%%%%%%%%%%%%%%%%%%%%%%%%%%%%%%%%%%%%%%%%%%%%%%%%%%%%%%%%%%%%%%%%%%%%%%%%%%%%%%%%%%%%%%%%%%%%%%%%%%
\textbf{\cite{weinstein2010characterization}} Western countries experience the Harmattan season, which occurs between November and March during the dry season. The dry and dusty northeasterly trade wind blows from the Saharan desert across west Africa into the Gulf of Guinea during this season. The Harmattan brings desert-like weather conditions that reduce humidity, raise daily temperatures, prevent rainfall formation, and produce large clouds of dust that cause dust storms or sandstorms. Tornadoes can form when the Harmattan interacts with monsoon winds. Moist equatorial air masses from the Atlantic and Pacific oceans bring annual monsoon rains from May to September. This wind blows from the south-west and brings rain between May and September (wet monsoon).

\begin{figure}[ht!]
    \centering
		\caption{Monthly PM2.5 concentrations in West Africa}
		% include second image
		\includegraphics[width=1.02\linewidth]{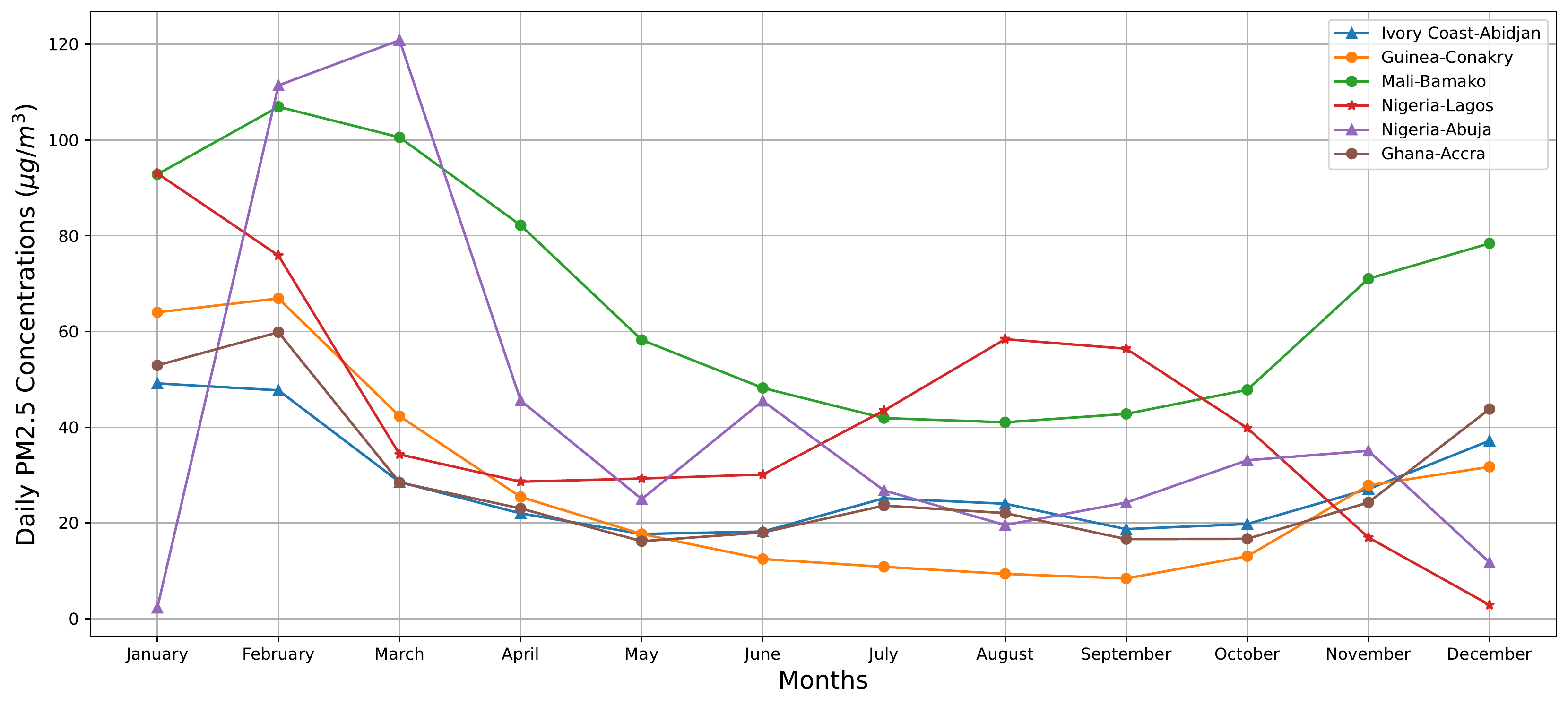}  
		\label{fig:sub-second_monthly}
\end{figure}

\begin{figure}[ht!]
    \centering
		\caption{Monthly PM2.5 concentrations in Central Africa}
		% include second image
		\includegraphics[width=1.02\linewidth]{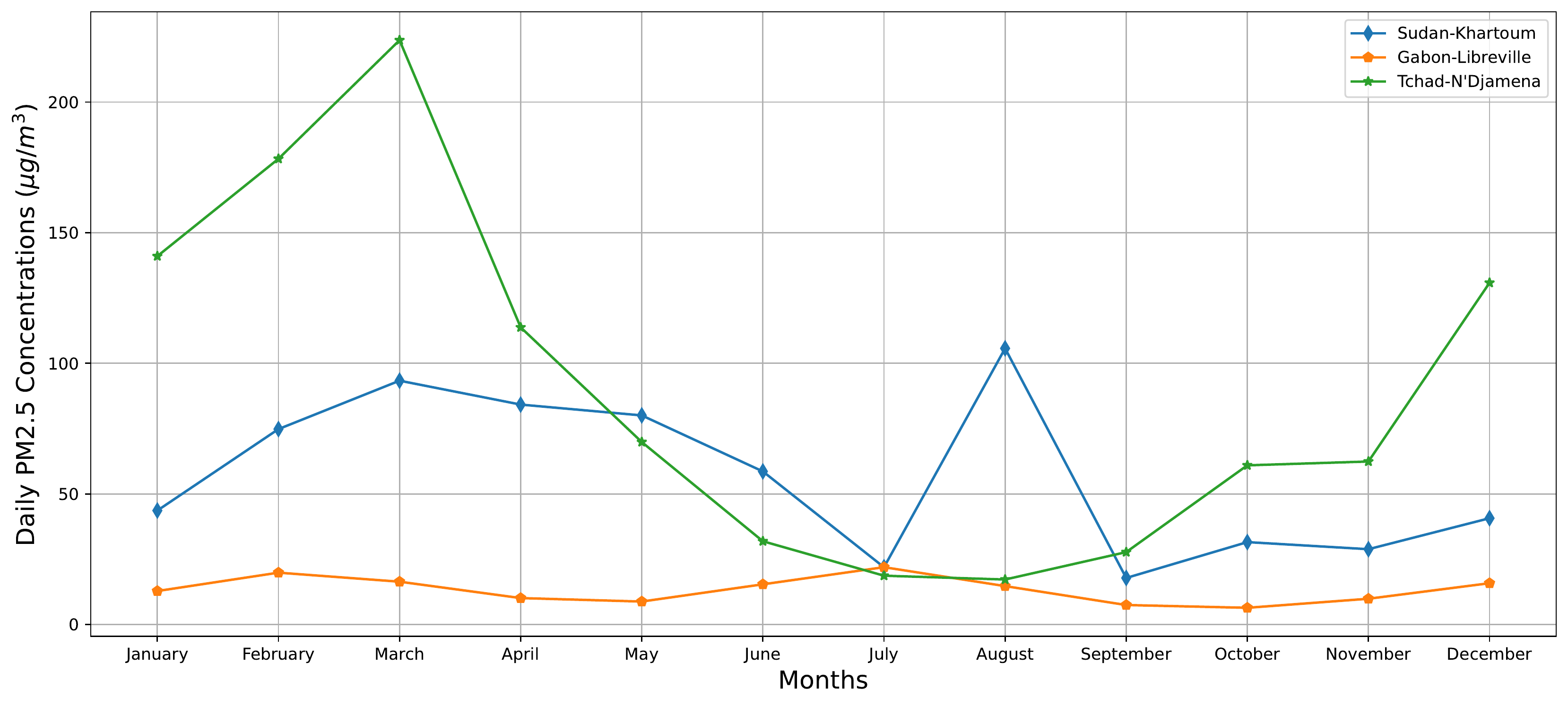}  
		\label{fig:sub-third_monthly}
\end{figure}

\begin{figure}[ht!]
    \centering
		\caption{Monthly PM2.5 concentrations in North Africa}
		% include second image
		\includegraphics[width=1.02\linewidth]{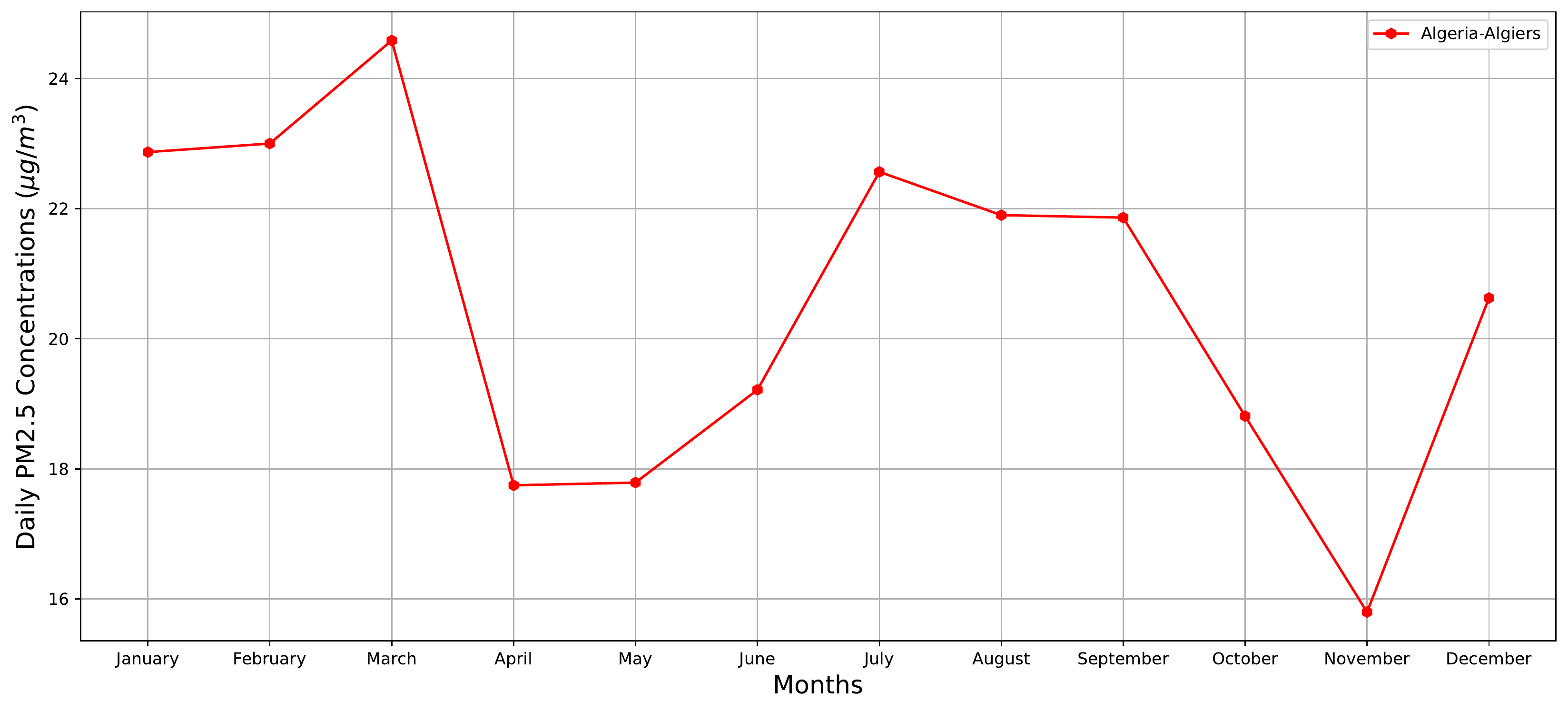}  
		\label{fig:sub-fourth_monthly}
\end{figure}

The seasonal variation of PM2.5 levels in West African cities is described in the figure \textbf{[\ref{fig:sub-second_monthly}]}. The decline of PM2.5 concentrations in the period of April to July is due to the rainy season across the region. In the rainy season, Conakry had the lowest monthly average PM2.5 concentrations of $\SI{8.40}{\micro\gram}/m^3$ and $\SI{9.37}{\micro\gram}/m^3$ in September and August, while in the dry season, Abuja, Bamako, and Lagos had the highest monthly average PM2.5 concentrations of $\SI{120.84}{\micro\gram}/m^3$, $\SI{106.92}{\micro\gram}/m^3$ and $\SI{93.02}{\micro\gram}/m^3$ in March, February, and January, respectively.
\FloatBarrier
%%%%%%%%%%%%%%%%%%%%%%%%%%%%%%%%%%%%%%%%%%%%%%%%%%%%%%%%%%%%%%%%%%%%%%%%%%%%%%%%%%%%%%%%%%%%%%%%%%%%%%%%%%%%%%%%%%%%%%%%%%%%%%%%%%%%%%%%%%%%%%%%%%
The monthly variations of PM2.5 levels in cities in Africa's central region are shown in figure \textbf{[\ref{fig:sub-third_monthly}]}. According to the figure, the highest levels of PM2.5 in N'Djamena were observed in March and February. After March, emissions decrease as the country enters the rainy season, which reduces the amount of dust in N'Djamena. The rate of emission in Libreville is lower than in Khartoum and N'Djamena, indicating that these two countries which are close to the Sahara desert have high monthly mean mass concentrations of PM2.5. Khartoum has a subtropical desert climate, with relative rainfall from the African monsoon from the south in July and September. High emissions occur in December and February, when the temperature is warm during the day and cool at night. This temperature variation, especially during the day, raises the rate of PM2.5 emissions in Khartoum.
\FloatBarrier
%%%%%%%%%%%%%%%%%%%%%%%%%%%%%%%%%%%%%%%%%%%%%%%%%%%%%%%%%%%%%%%%%%%%%%
The seasonal spatio-temporal variation of PM2.5 in Algiers is represented in figure \textbf{[\ref{fig:sub-fourth_monthly}]}. The monthly average value of PM2.5 ratios falls month after month. Many sandstorms hit Algiers in January, February, and March, raising PM2.5 levels. The monthly mean mass concentrations of PM2.5 in February, January, and March were $\SI{22.86}{\micro\gram}/m^3$, $\SI{23.0005}{\micro\gram}/m^3$, and $\SI{24.58}{\micro\gram}/m^3$, respectively. PM2.5 levels have risen in the months of May, June, and July with monthly average of $\SI{}{\micro\gram}/m^3$, $\SI{}{\micro\gram}/m^3$, and $\SI{}{\micro\gram}/m^3$, respectively because the weather is warm. The lowest monthly mean PM2.5 concentration of $\SI{15.79}{\micro\gram}/m^3$ was observed in November, as the country entered the winter season.
\FloatBarrier
%%%%%%%%%%%%%%%%%%%%%%%%%%%%%%%%%%%%%%%%%%%%%%%%%%%%%%%%%%%%%%%%%%%%%%%%%%%%%%%%%%%%%%%%%%%%%%%%%%%%%%%%%%%%%%%%%%%%%%%%%%%%%%%%%%%%%%%%%%%%%%%%%%%%%%%%%%%%%%%%%%%%%%%%%%%%%%%%%%%%
\subsection{Forecasting}
\begin{figure}[ht!]
    \centering
		\caption{Forecasting with ARIMA, Neural Networks, and Gaussian Processes on Kigali PM2.5 data}
		% include second image
		\includegraphics[width=1.02\linewidth]{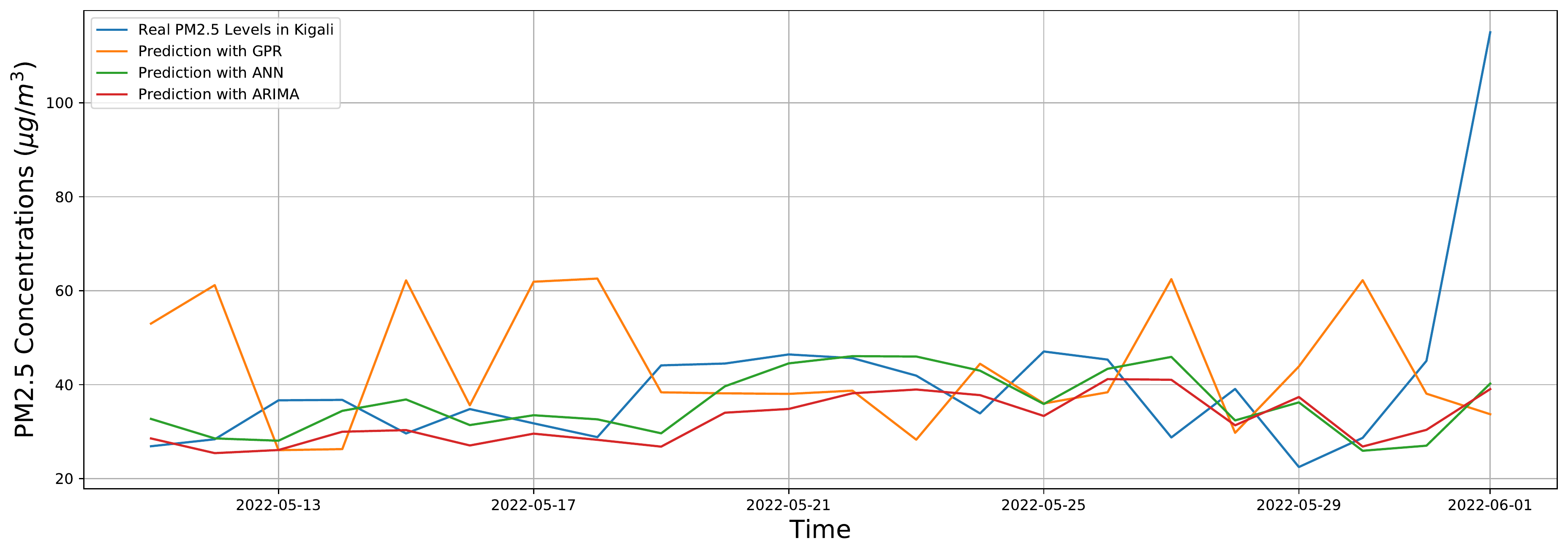}  
		\label{model1}
\end{figure}

\begin{figure}[ht!]
    \centering
		\caption{Forecasting with ARIMA, Neural Networks, and Gaussian Processes on Conakry PM2.5 data}
		% include second image
		\includegraphics[width=1.02\linewidth]{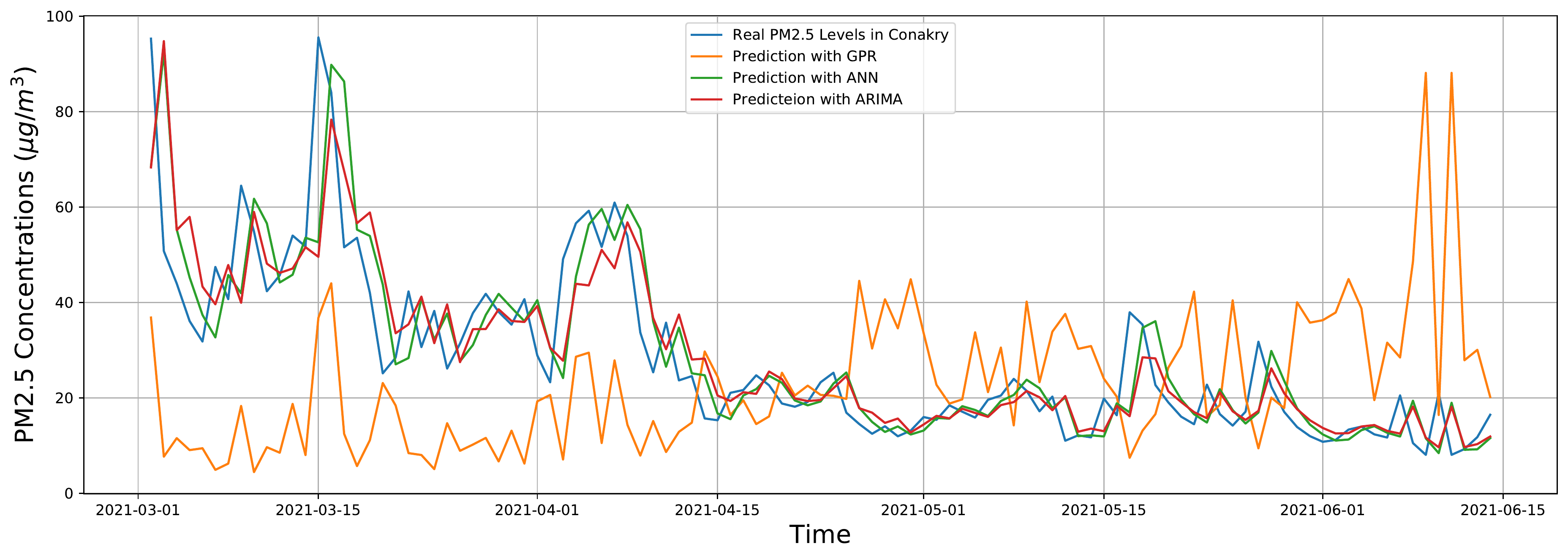}  
		\label{model2}
\end{figure}

\begin{figure}[ht!]
    \centering
		\caption{Forecasting with ARIMA, Neural Networks, and Gaussian Processes on Bamako PM2.5 data}
		% include second image
		\includegraphics[width=1.02\linewidth]{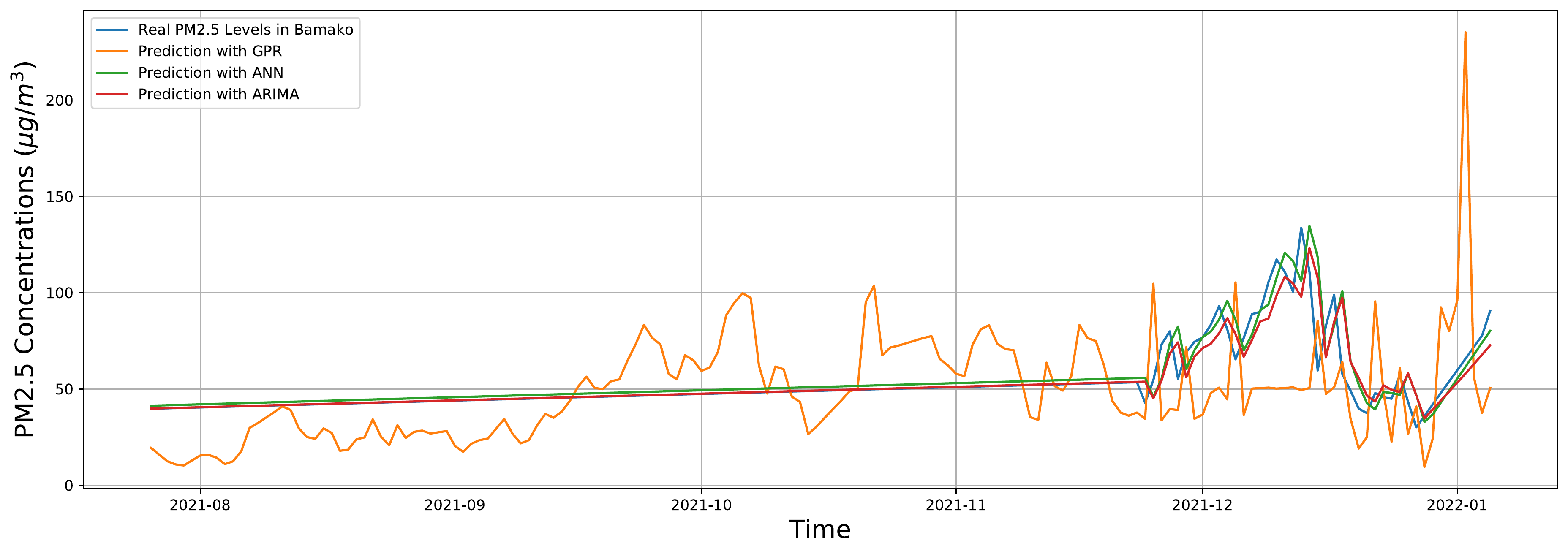}  
		\label{model3}
\end{figure}

\begin{figure}[ht!]
    \centering
		\caption{Forecasting with ARIMA, Neural Networks, and Gaussian Processes on Lagos PM2.5 data}
		% include second image
		\includegraphics[width=1.02\linewidth]{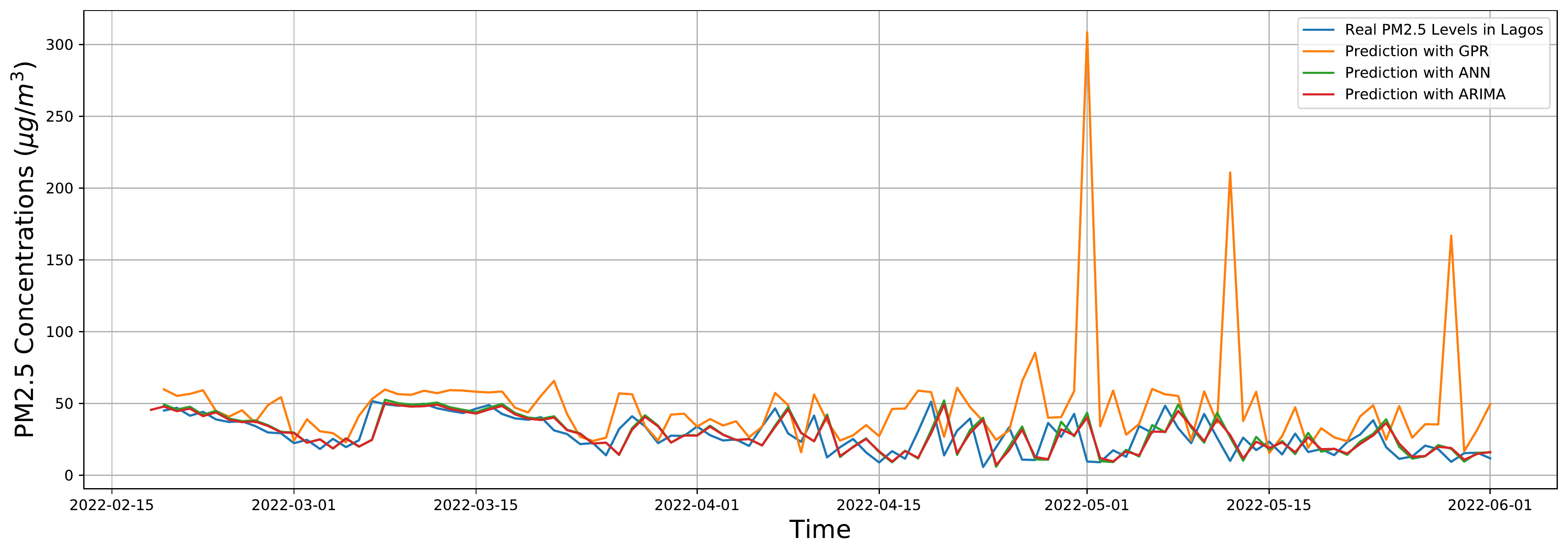}  
		\label{model33}
\end{figure}

\begin{figure}[ht!]
    \centering
		\caption{Forecasting with ARIMA, Neural Networks, and Gaussian Processes on Abuja PM2.5 data}
		% include second image
		\includegraphics[width=1.02\linewidth]{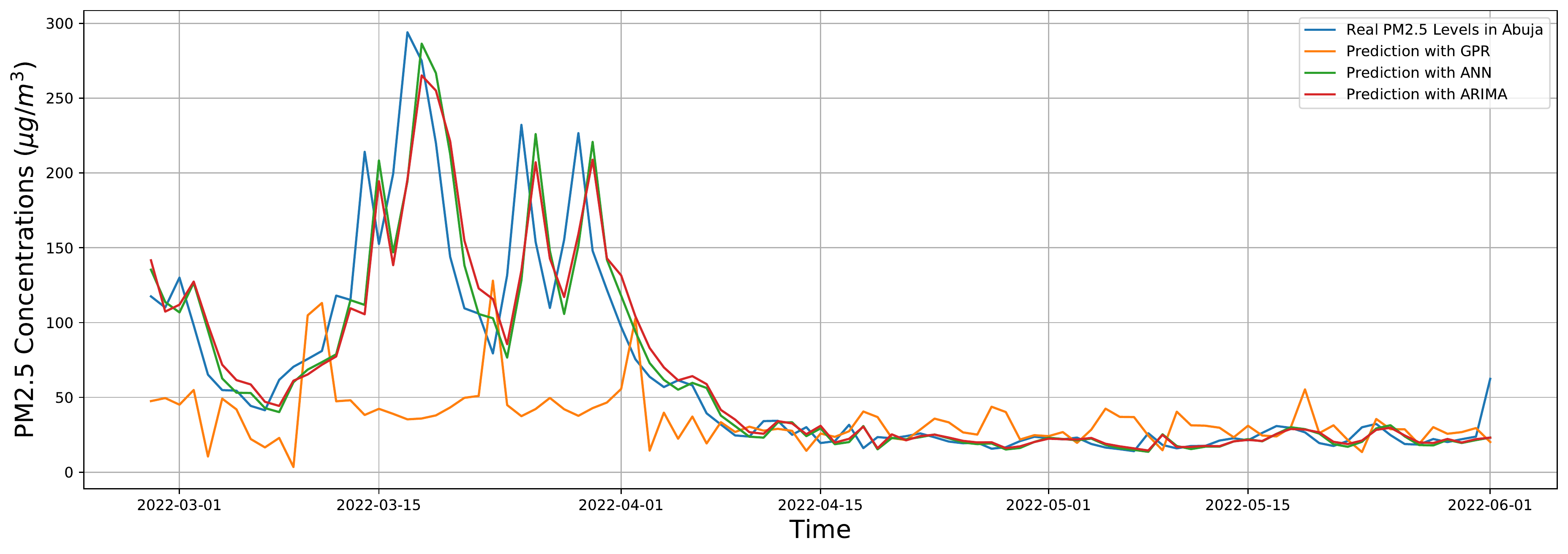}  
		\label{model4}
\end{figure}

\begin{figure}[ht!]
    \centering
		\caption{Forecasting with ARIMA, Neural Networks, and Gaussian Processes on Libreville PM2.5 data}
		% include second image
		\includegraphics[width=1.02\linewidth]{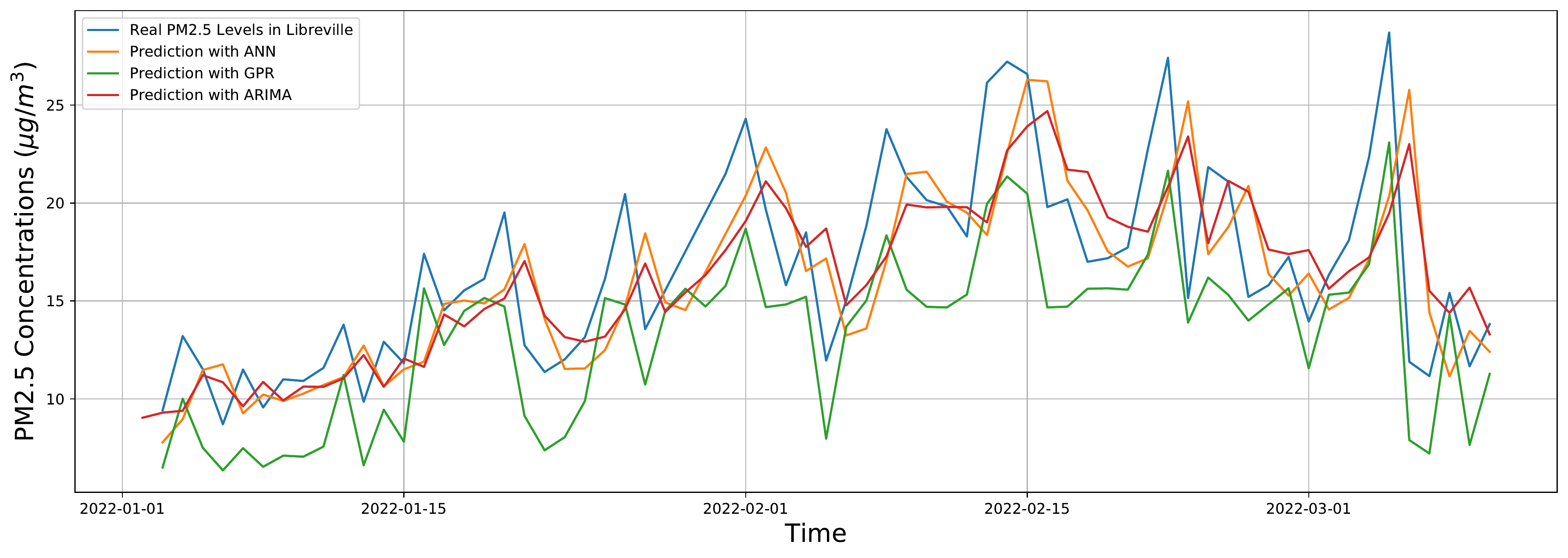}  
		\label{model5}
\end{figure}

\begin{figure}[ht!]
    \centering
		\caption{Forecasting with ARIMA, Neural Networks, and Gaussian Processes on Algiers PM2.5 data}
		% include second image
		\includegraphics[width=1.02\linewidth]{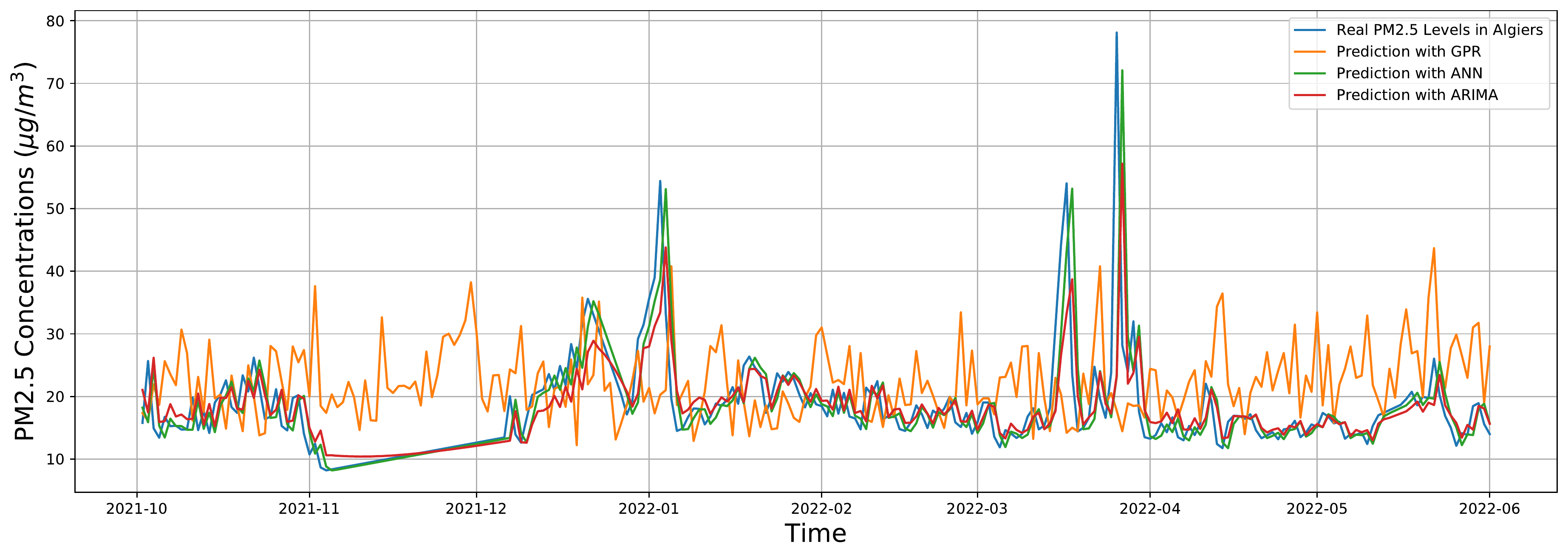}  
		\label{model6}
\end{figure}

\begin{figure}[ht!]
    \centering
		\caption{Forecasting with ARIMA, Neural Networks, and Gaussian Processes on Accra PM2.5 data}
		% include second image
		\includegraphics[width=1.02\linewidth]{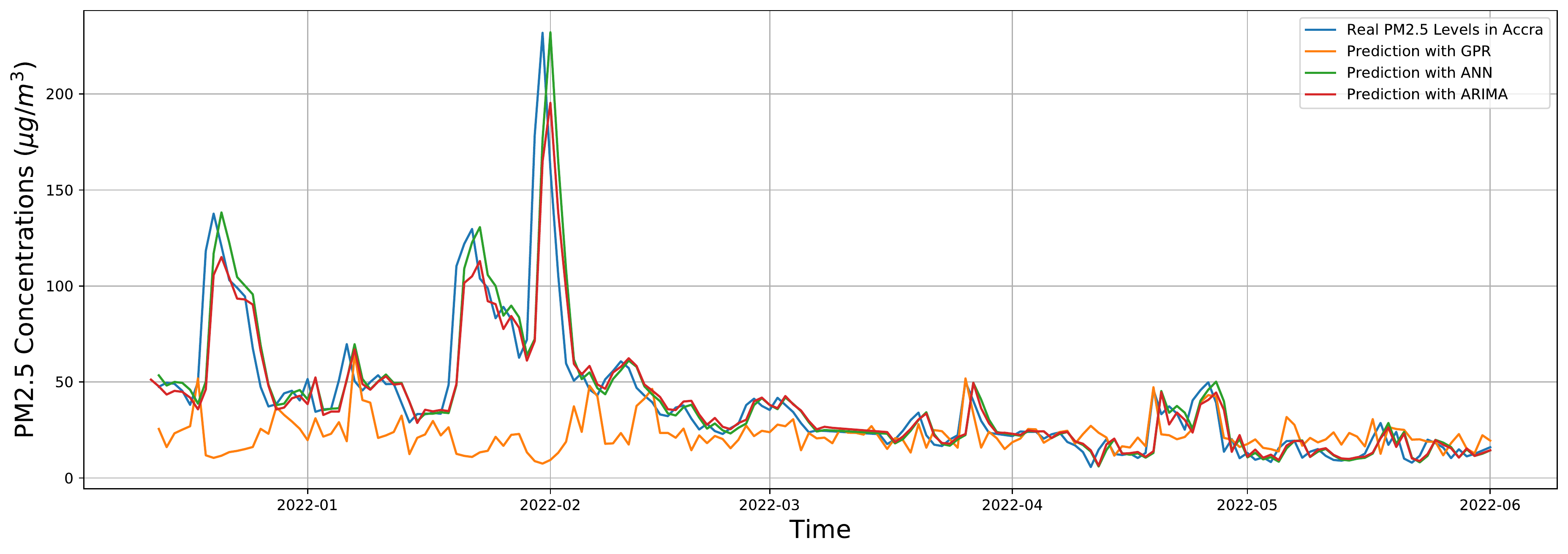}  
		\label{model7}
\end{figure}

\begin{figure}[ht!]
    \centering
		\caption{Forecasting with ARIMA, Neural Networks, and Gaussian Processes on Kinshasa PM2.5 data}
		% include second image
		\includegraphics[width=1.02\linewidth]{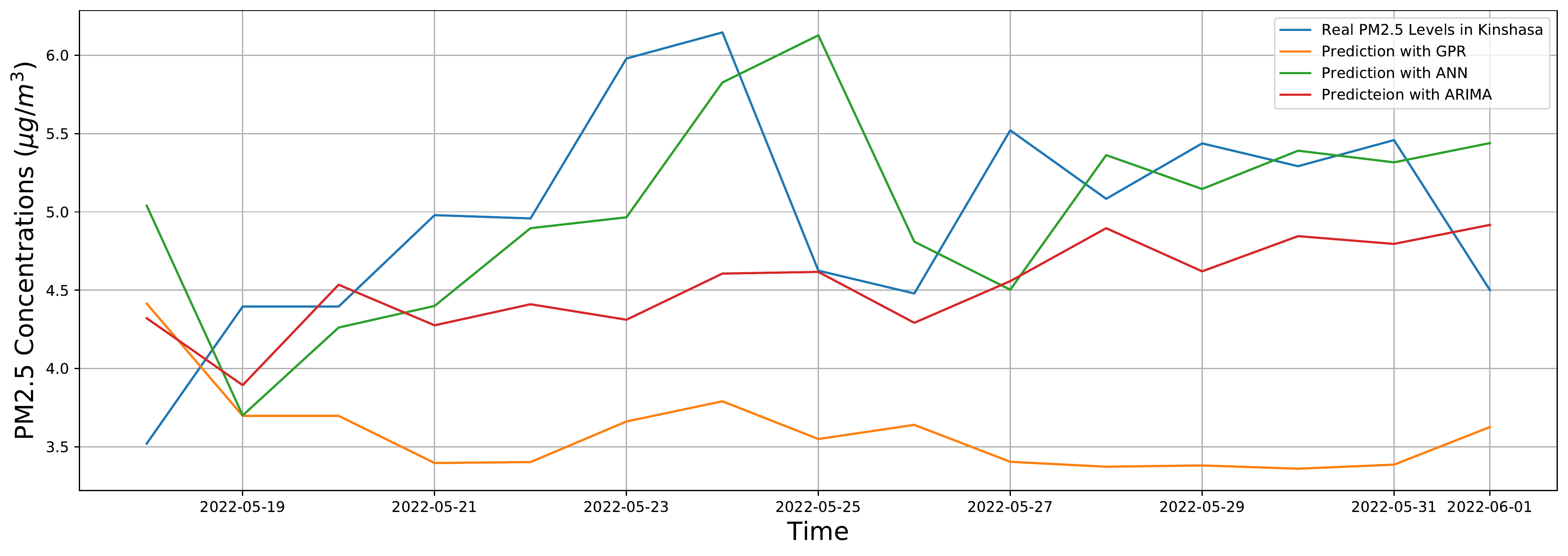}  
		\label{model8}
\end{figure}

\begin{figure}[ht!]
    \centering
		\caption{Forecasting with ARIMA, Neural Networks, and Gaussian Processes on Kampala PM2.5 data}
		% include second image
		\includegraphics[width=1.02\linewidth]{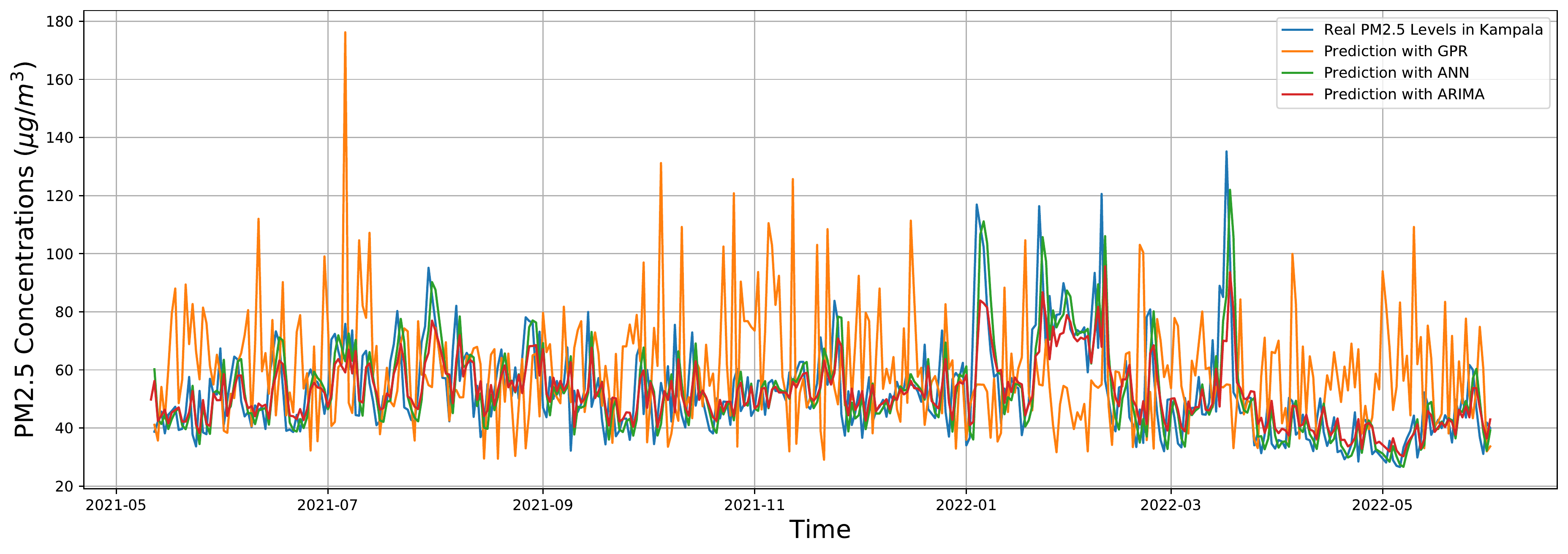}  
		\label{model9}
\end{figure}

\begin{figure}[ht!]
    \centering
		\caption{Forecasting with ARIMA, Neural Networks, and Gaussian Processes on Nairobi PM2.5 data}
		% include second image
		\includegraphics[width=1.02\linewidth]{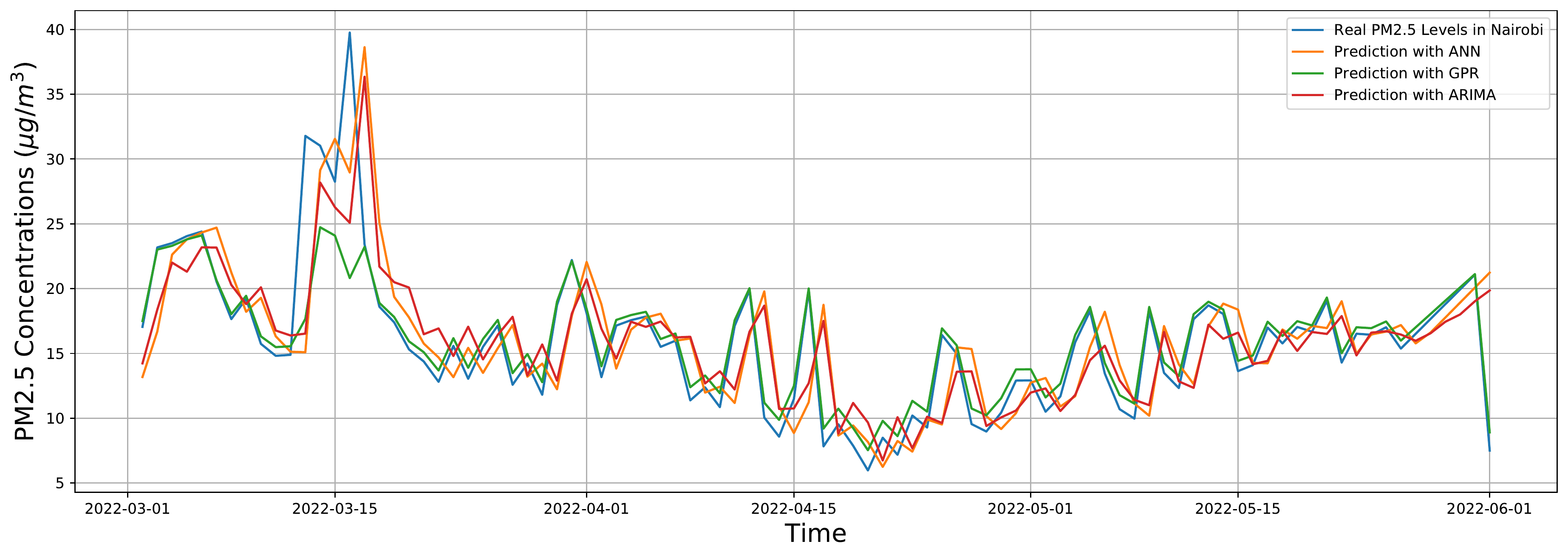}  
		\label{model10}
\end{figure}

\begin{figure}[ht!]
    \centering
		\caption{Forecasting with ARIMA, Neural Networks, and Gaussian Processes on N'Djamena PM2.5 data}
		% include second image
		\includegraphics[width=1.02\linewidth]{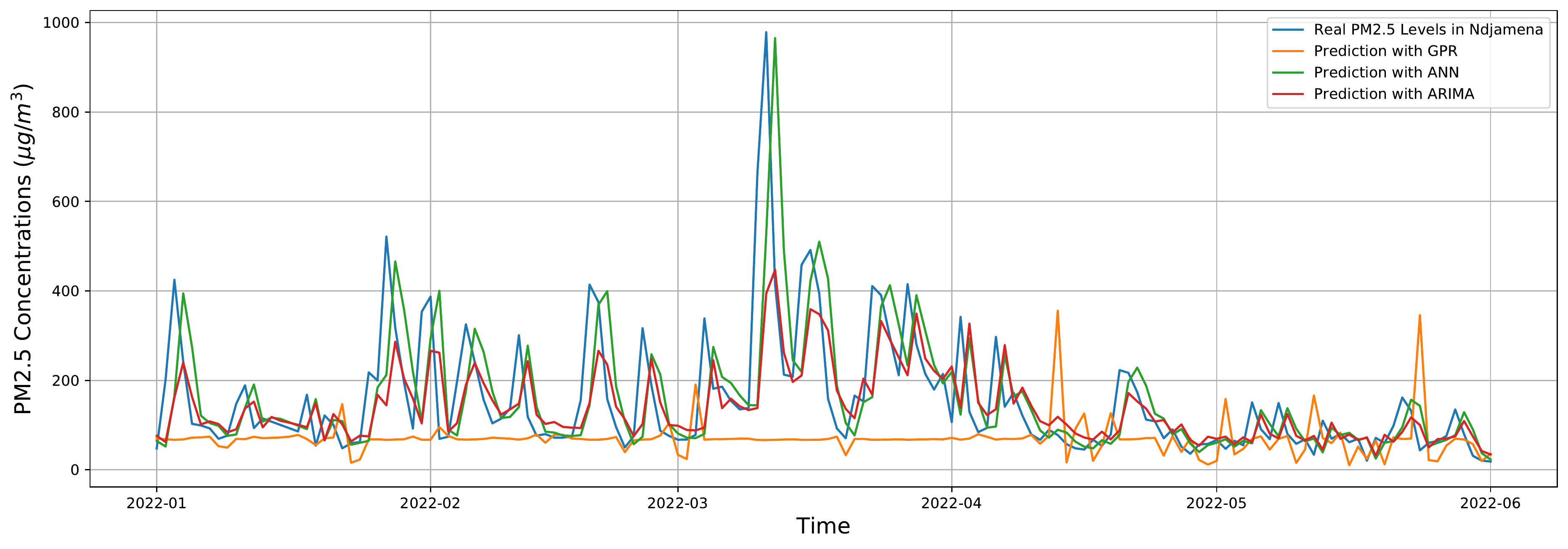}  
		\label{model11}
\end{figure}

\begin{figure}[ht!]
    \centering
		\caption{Forecasting with ARIMA, Neural Networks, and Gaussian Processes on Khartoum PM2.5 data}
		% include second image
		\includegraphics[width=1.02\linewidth]{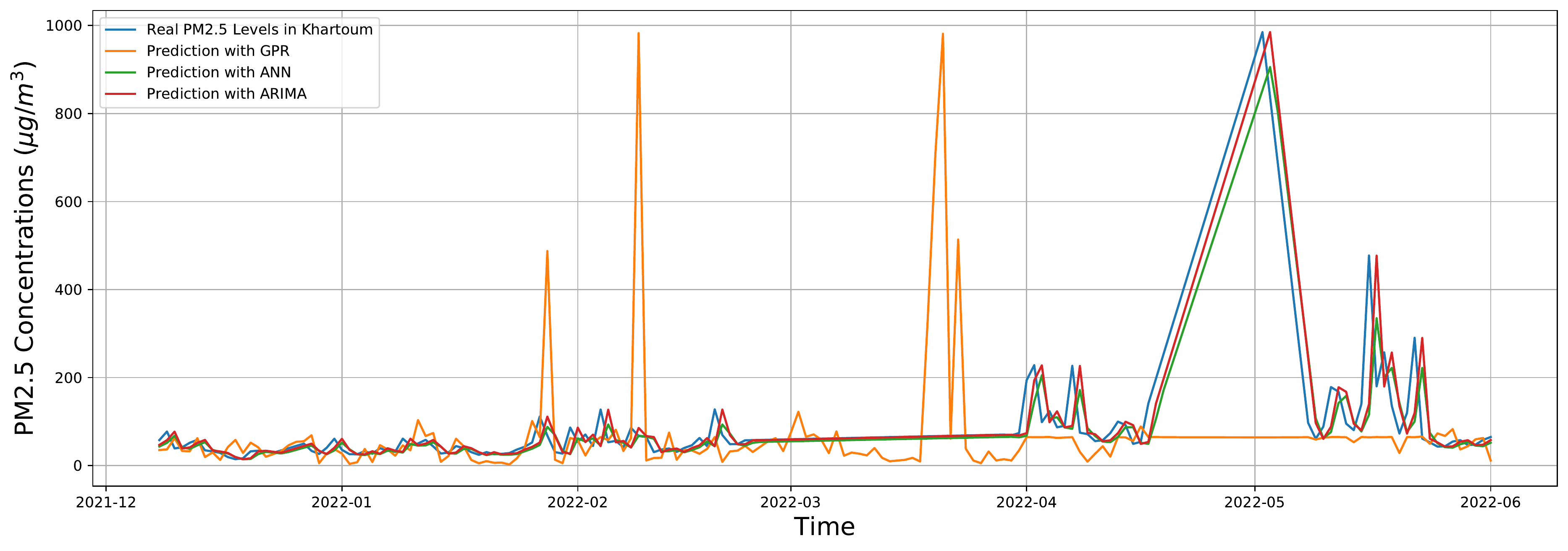}  
		\label{model12}
\end{figure}

\begin{figure}[ht!]
    \centering
		\caption{Forecasting with ARIMA, Neural Networks, and Gaussian Processes on Addis Ababa PM2.5 data}
		% include second image
		\includegraphics[width=1.02\linewidth]{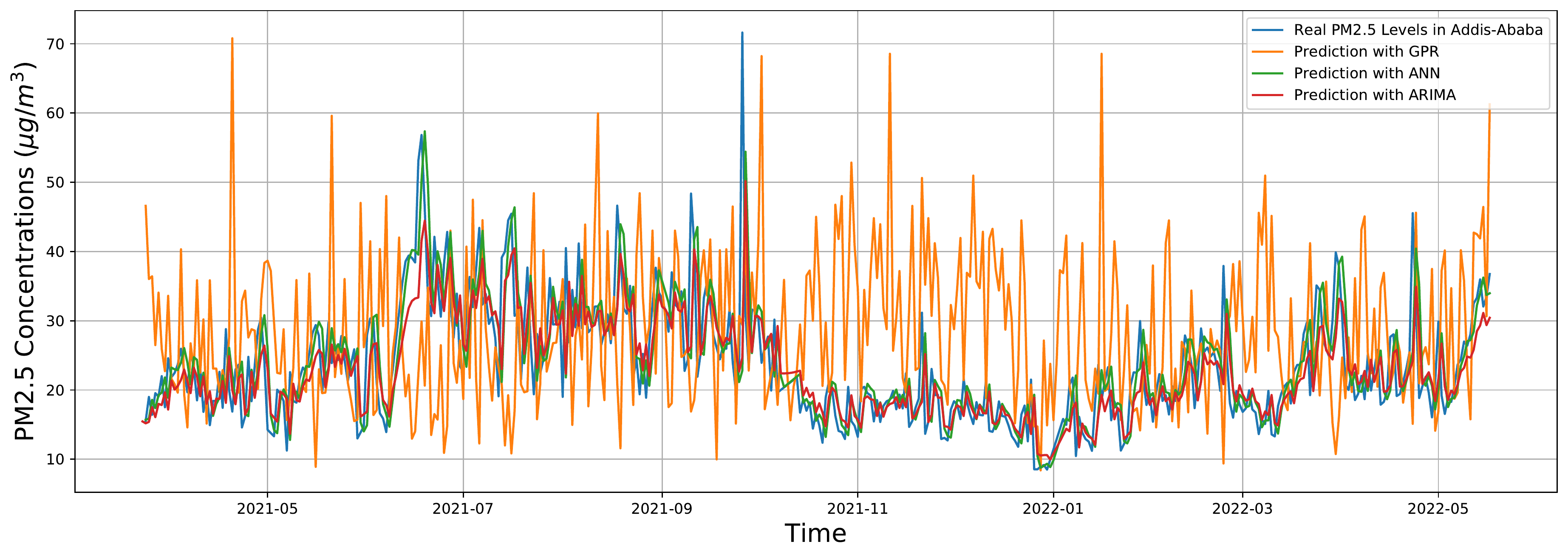}  
		\label{model13}
\end{figure}

\begin{figure}[ht!]
\centering
\caption{Forecasting with ARIMA, Neural Networks, and Gaussian Processes on Antananarivo PM2.5 data}
	% include second image
\includegraphics[width=1.02\linewidth]{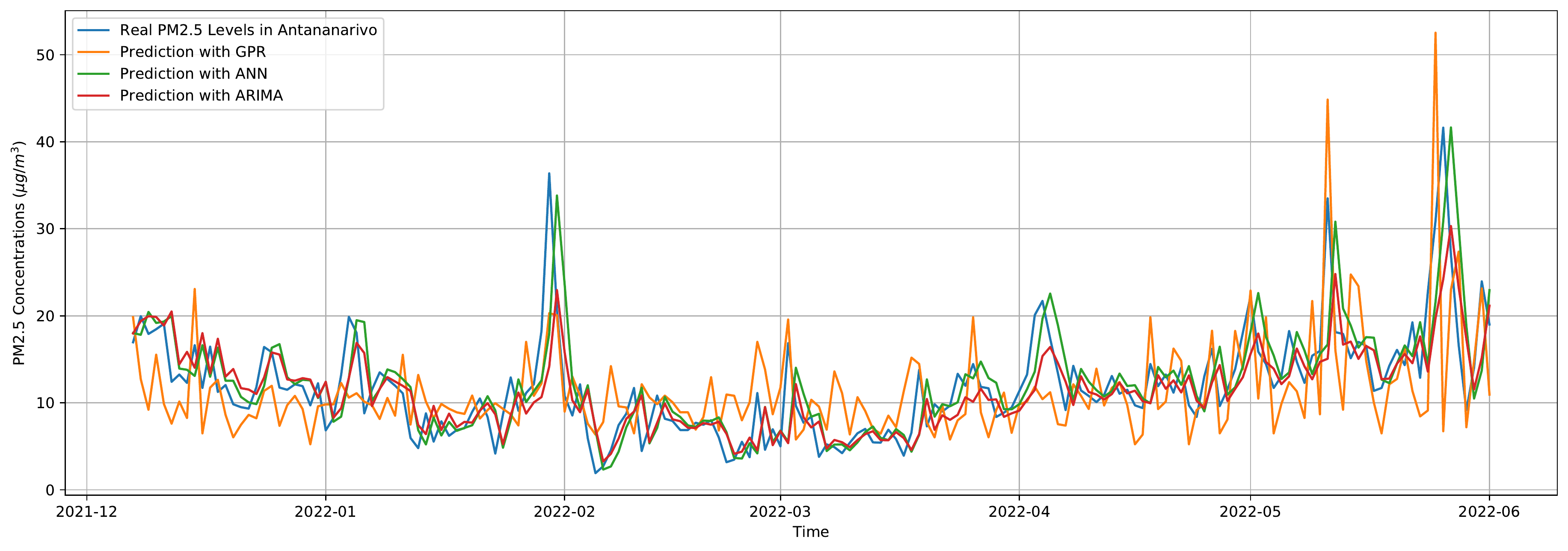} 
\label{model14}
\end{figure}

\begin{figure}[ht!]
    \centering
		\caption{Forecasting with ARIMA, Neural Networks, and Gaussian Processes on Abidjan PM2.5 data}
		% include second image
		\includegraphics[width=1.02\linewidth]{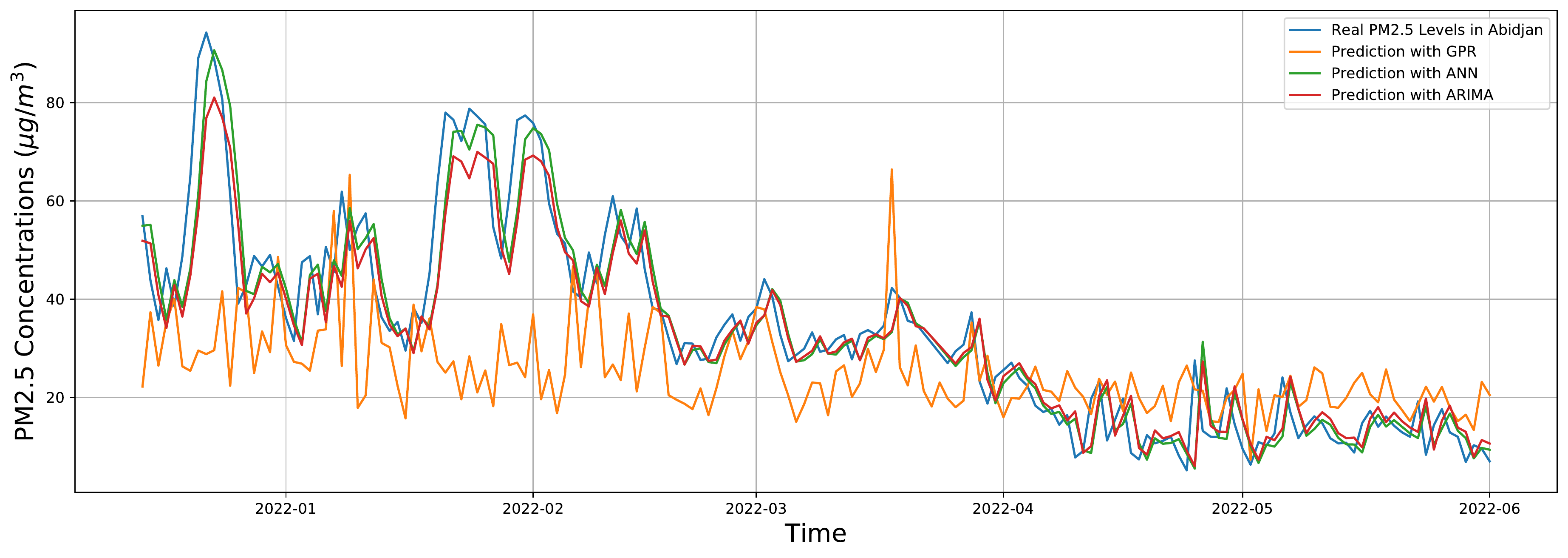}  
		\label{model15}
\end{figure}

The figures from \textbf{[\ref{model1} - \ref{model15}]} are for forecasting trends in real-time PM2.5 data in all of the cities studied. To train models, data was divided into two sets: training and testing. Both models were used to forecast PM2.5 levels in various African cities.
\FloatBarrier
%%%%%%%%%%%%%%%%%%%%%%%%%%%%%%%%%%%%%%%%%%%%%%%%%%%%%%%%%%%%%%%%%%%%%%%%%%%%%%%%%%%%%%%%%%%%%%%%%
\FloatBarrier

%%%%%%%%%%%%%%%%%%%%%%%%%%%%%%%%%%%%%%%
\FloatBarrier
\FloatBarrier
%%%%%%%%%%%%%%%%%%%%%%%%%%%%%%%%%%%%%%%%%%%%%%%%%%%%%%%%%%%%%%%%%%%%%%%%%%%%%%%%%%%%%%%%%%%%%%%%%%
\FloatBarrier
\FloatBarrier
%%%%%%%%%%%%%%%%%%%%%%%%%%%%%%%%%%%%%%%%%%%%%%%%%%%%%%%%%%%%%%%%%%%%%%%%%%%%%%%%%%%%%%%%%%%%%%%%%

\FloatBarrier
\FloatBarrier
%%%%%%%%%%%%%%%%%%%%%%%%%%%%%%%%%%%%%%%%%%%%%%%%%%%%%%%%%%%%%%%%%%%%%%%%%%%%%%%%%%%%%%%%%%%%%%%%%%

\FloatBarrier
\FloatBarrier
%%%%%%%%%%%%%%%%%%%%%%%%%%%%%%%%%%%%%%%%%%%%%%%%%%%%%%%%%%%%%%%%%%%%%%%%%%%%%%%%%%%%%%%%%%%%%%%%%%

\FloatBarrier
\FloatBarrier
%%%%%%%%%%%%%%%%%%%%%%%%%%%%%%%%%%%%%%%%%%%%%%%%%%%%%%%%%%%%%%%%%%%%%%%%%%%%%%%%%%%%%%%%%%%%%%%%%%

\FloatBarrier
\FloatBarrier
%%%%%%%%%%%%%%%%%%%%%%%%%%%%%%%%%%%%%%%%%%%%%%%%%%%%%%%%%%%%%%%%%%%%%%%%%%%%%%%%%%%%%%%%%%%%%%%%%%
\FloatBarrier
\FloatBarrier
%%%%%%%%%%%%%%%%%%%%%%%%%%%%%%%%%%%%%%%%%%%%%%%%%%%%%%%%%%%%%%%%%%%%%%%%%%%%%%%%%%%%%%%%%%%%%%%%%%

\FloatBarrier
\FloatBarrier
%%%%%%%%%%%%%%%%%%%%%%%%%%%%%%%%%%%%%%%%%%%%%%%%%%%%%%%%%%%%%%%%%%%%%%%%%%%%%%%%%%%%%%%%%%%%%%%%%%

\FloatBarrier
\FloatBarrier
%%%%%%%%%%%%%%%%%%%%%%%%%%%%%%%%%%%%%%%%%%%%%%%%%%%%%%%%%%%%%%%%%%%%%%%%%%%%%%%%%%%%%%%%%%%%%%%%%%

\FloatBarrier
\FloatBarrier
%%%%%%%%%%%%%%%%%%%%%%%%%%%%%%%%%%%%%%%%%%%%%%%%%%%%%%%%%%%%%%%%%%%%%%%%%%%%%%%%%%%%%%%%%%%%%%%%%%

\FloatBarrier
\FloatBarrier
%%%%%%%%%%%%%%%%%%%%%%%%%%%%%%%%%%%%%%%%%%%%%%%%%%%%%%%%%%%%%%%%%%%%%%%%%%%%%%%%%%%%%%%%%%%%%%%%%%

\FloatBarrier
\FloatBarrier
%%%%%%%%%%%%%%%%%%%%%%%%%%%%%%%%%%%%%%%%%%%%%%%%%%%%%%%%%%%%%%%%%%%%%%%%%%%%%%%%%%%%%%%%%%%%%%%%%

\FloatBarrier
\FloatBarrier
%%%%%%%%%%%%%%%%%%%%%%%%%%%%%%%%%%%%%%%%%%%%%%%%%%%%%%%%%%%%%%%%%%%%%%%%%%%%%%%%%%%%%%%%%%%%%%%%%%

\FloatBarrier
\FloatBarrier
%%%%%%%%%%%%%%%%%%%%%%%%%%%%%%%%%%%%%%%%%%%%%%%%%%%%%%%%%%%%%%%%%%%%%%%%%%%%%%%%%%%%%%%%%%%%%%%%%%%%%%%%%%%%%%%%%%%%%%%%%%%%%%%%%%%%%%%%%%%%%%%%%%%%%%%%%%%%%%%%%%%%%%%%%%%%%%%%%%%%%%%%%%%%%%%%%%%%%%%%%
\section{Model Comparison}
The prediction of air pollution, particularly particulate matter less than 2.5 microns in diameter, is location dependent because the sources of emissions differ from city to city and country to country. The forecasting results show that neural network and ARIMA models outperformed Gaussian Process Regression algorithms in adapting to real-time PM2.5 trends. The models were compared using three statistical metrics: root mean square, mean absolute error, and mean absolute percentage error to determine the accuracy of the model in adapting to actual PM2.5 levels. The table \textbf{[\ref{tab:my_label}]} compares the performance metrics of ARIMA, Neural Networks, and Gaussian Processes.
\begin{table}[ht!]
\caption{The Root Mean Square Errors (RMSEs) of ARIMA, Neural Networks, and Gaussian Processes.}
	\centering
	\resizebox{\columnwidth}{!}{%
	\begin{tabular}{ p{1.82cm}p{0.89cm}p{0.89cm}p{0.89cm}  }
% 		\hline
% 		\hline
\toprule
		\multicolumn{1}{c}{\textbf{US Embassy in cities}} &\multicolumn{3}{c}{\textbf{RMSE}}\\
% 		\hline
\midrule
		& \textbf{{\footnotesize ARIMA}} &\textbf{{\footnotesize AN N}}&\textbf{{\footnotesize GPR}}\\
% 		\hline
\midrule
	  Kigali  & 18.441  &17.770  &26.295   \\
% 		\hline
		Conakry & 10.544   & 11.386   &27.336 \\
% 		\hline
		Bamako &7.989 &8.138 &29.383  \\
% 		\hline
		Lagos   &11.494 &11.932 &42.882    \\
% 		\hline
		Abuja & 28.168    &29.253 &67.153  \\
% 		\hline
		Libreville &3.756 &4.244    &3.888  \\
% 		\hline
		Algiers &5.785  &6.260  &10.954  \\
% 		\hline
		Accra &15.328   &16.672  &39.869  \\
% 		\hline
		Kinshasa& 0.789   &0.773  &1.631   \\
% 		\hline
		Kampala&12.179    &13.056  &27.059    \\
% 		\hline
		Nairobi&4.265    & 4.480 &2.692   \\
% 		\hline
		N'Djamena&108.71   &123.67  &165.84  \\
% 		\hline
		Khartoum& 60.488   &60.170  &231.69    \\
% 		\hline
		Addis Ababa& 5.801   &6.237  & 15.368  \\
% 		\hline
		Antananarivo& 4.282   &4.534  &5.664     \\
% 		\hline
		Abidjan&  7.660  &7.594  &20.509     \\
% 		\hline
% 		\hline
\bottomrule
	\end{tabular}
}
	\label{tab:my_label}
\end{table}
\FloatBarrier
\vspace{0.9mm}
According to the RMSE results, both the ARIMA and the Neural Network outperform the Gaussian process. This means that neural networks, as a machine learning approach, can be used to analyze and predict time series data, as well as to solve real-world problems like air pollution. Although the results of the Gaussian process regression are not as precise as those of the other models, the study shows that they can be used to forecast air quality data.
%%%%%%%%%%%%%%%%%%%%%%%%%%%%%%%%%%%%%%%%%%%%%%%%%%%%%%%%%%%%%%%%%%%%%%%%%%%%%%%%%%%%%%%%%%%%%%%%%%%%%%%%%%%%%%%%%%%%%%%%%%%%%%%%%%%%%%%%%%%%%%%%%%%%%%%%%%%%%%%%%%%%%%%%%%%%%%%%%%%%%%%%%%%%
% \multirow{2}{*}{Alg.}
\section{Conclusion}
This study has focused on the analysis of PM2.5 variations across Africa. As it has been demonstrated in this paper, data-driven models such as Auto-Rregressive Integrated Moving Average, Neural Network, and Gaussian Process Regression can be used to forecast PM2.5 trends in Africa. Future work should consider other air pollutants that affect human health, such as PM10, SO2, NOx, CO, and O3.
%%%%%%%%%%%%%%%%%%%%%%%%%%%%%%%%%%%%%%%%%%%%%%%%%%%%%%%%%%%%%%%%%%%%%%%%%%%%%%%%%%%%%%%%%%%%%%%%%%%%%%%%%%%%%%%%%%%%%%%%%%%%%%%%%%%%%%%%%%%%%%%%%%%%%%%%%%%%%%%%%%%%%%%%%%%%%%%%%%%%%%%%%%%%%%%%%%%%%%%%%%%%%%%%%%%%%%%%%%%%%%%%%%%%%%%%%%%%%%%%%%%%%%%%%%%%%%%%%%%%%%%%%%%%%%%%%%%%%%%%%%%%%%%%%%%%%%%%%%%%%%%%%%%%%%%%%%
 \section{Acknowledgments}
 The authors, P.G. and J-R.K., would like to thank the African Institute for Mathematical Sciences for their assistance, as well as the financial support provided by the Canadian government through Global Affairs Canada and the International Development Research Centre.
 %%%%%%%%%%%%%%%%%%%%%%%%%%%%%%%%%%%%%%%%%%%%%%%%%%%%%%%%%%%%%%%%%%%%%%%%%%%%%%%%%%%%%%%%%%%%%%%%%%%%%%%%%%%%%%%%%%%%%%%%%%%%%%%%%%%%%%%%%%%%%%%%%%%%%%%%%%%%%%%%%%%%%%%%%%%%%%%%%%%%%%%%%%%%%%%%%%%%%%%%%%%%%%%%%%%%%%%%%%%%%%%%%%%%%%%%%%%%%%%%%%%%%%%%%%%%%%%%%%%%%%%%%%%%%%%%%%%%%%%%%%%%%%%%%%%%%%%%%%%%%%%%%%%%%%
%\newpage

%\nocite{*} % This command displays all refs in the bib file. PLEASE DELETE IT BEFORE YOU SUBMIT YOUR MANUSCRIPT!
\bibliographystyle{abbrv}
\bibliography{references}

%%%%%%%%%%%%%%%%%%%%%%%%%%%%%%%%%%%%%%%%%%%%%%%%%%%%%%%%%%%%
%%% APPENDICES
%%%%%%%%%%%%%%%%%%%%%%%%%%%%%%%%%%%%%%%%%%%%%%%%%%%%%%%%%%%%

% \appendix
% \begin{appendixbox}
% % 
% \end{appendixbox}

%%%%%%%%%%%%%%%%%%%%%%%%%%%%%%%%%%%%%%%%%%%%%%%%%%%%%%%%%%%%%%%%%%%%%%%%%%%%%%%

\end{document}